\documentclass[aps, pre, onecolumn, amsmath, superscriptaddress]{revtex4}

\usepackage{fancyhdr}
\usepackage{calc}
\usepackage{amssymb}
\usepackage{setspace}
\usepackage{amsfonts}
\usepackage[innercaption]{sidecap}
\usepackage{graphicx}
\usepackage{graphicx,epstopdf}
\usepackage{xcolor, soul}
\usepackage{dcolumn}
\usepackage{bm}
\usepackage{mathrsfs} 
\usepackage{amsmath} 
\usepackage[colorlinks=true,linkcolor=blue,citecolor=blue]{hyperref}%
\usepackage{fontenc}
\usepackage{float}
\usepackage{amsthm}
\usepackage{subfigure}
\usepackage{color}
\usepackage{ragged2e}
\usepackage{enumerate}
\usepackage{hyperref}

\begin{document}	

	\newcommand{\R} 
	{
		\mathbb R
	}
	
	\newcommand{\M} 
	{
		\mathbb M
	}
	
	\newcommand{\A} 
	{
		\mathcal A
	}
    \title{Optimal test-kit based intervention strategy  of epidemic spreading in heterogeneous complex networks}
    \author{Subrata Ghosh}
    	\thanks{Equal contribution}
    \affiliation{Physics and Applied Mathematics Unit, Indian Statistical Institute, 203 B. T. Road, Kolkata 700108, India}
    \author{Abhishek Senapati}
    	\thanks{Equal contribution}
       \affiliation{Agricultural and Ecological Research Unit, Indian Statistical Institute, 203 B. T. Road, Kolkata 700108, India}
 \author{Joydev Chattopadhyay}
\affiliation{Agricultural and Ecological Research Unit, Indian Statistical Institute, 203 B. T. Road, Kolkata 700108, India}   
     	
 \author{Chittaranjan Hens}
\thanks{Corresponding Author}
\email{chittaranjanhens@gmail.com}
\affiliation{Physics and Applied Mathematics Unit, Indian Statistical Institute, 203 B. T. Road, Kolkata 700108, India}

	\author{Dibakar Ghosh}
\affiliation{Physics and Applied Mathematics Unit, Indian Statistical Institute, 203 B. T. Road, Kolkata 700108, India}



    \begin{abstract}
We propose a  deterministic compartmental model of infectious disease which considers the test-kits as an important ingredient for the  suppression and mitigation of epidemics. 
A rigorous simulation (with  analytical argument) is provided to reveal the effective reduction  of final outbreak size and peak of infection as a function of basic reproduction number in  a single patch. Further, to study the impact of  long and short-distance human migration among the patches, we have  considered   heterogeneous networks where the linear diffusive connectivity is determined by the network link structure. We  numerically confirm that implementation of test-kits in the fraction of nodes (patches) having larger degrees or betweenness centralities can  reduce the peak of infection (as well as final outbreak size) significantly. A next-generation matrix based analytical treatment is provided to find out the critical transmission probability in the entire network for the onset of epidemics. Finally, the optimal intervention strategy is validated in two real networks: global airport networks and transportation networks of Kolkata, India.
\end{abstract}
\maketitle
\section{Introduction}
In the last two decades, human populations have experienced the flare-up of diverse infectious diseases. For instance, outbreak of 'Severe Acute Respiratory Syndrome' (SARS) in 2003 \cite{colizza2007predictability,hufnagel2004forecast}, swine flu pandemic in 2009 \cite{fraser2009pandemic}, and more recently, the ongoing COVID-19 pandemic \cite{zhang2020changes}, all have large-scale impact on the public health as well as the socio-economic condition of the affected country.
In the absence of effective vaccine interventions,  to fight against such drastic pandemics,  diverse controlled  interventions strategies are required to mitigate   the  spread of disease and reduce the outbreak size \cite{zhang2020changes,ferguson2006strategies,pastor2015epidemic}.   {These strategies  are broadly involved with the implementation of  strict lock down in social communities  e.g. travel restriction, shutdown of non-essential services. 
\par
In order to gain deeper insights on the disease transmission mechanism and to analyze the efficiency of intervention strategies,  the researchers utilize the  compartmental mathematical models ranging from  stochastic (Markovian or non-Markovian) to  deterministic frameworks~ \cite{arenas2020mathematical,pastor2015epidemic}. In these  set-ups, the severity of the disease is determined by one of the key parameters of the system: the basic reproduction number (${\mathcal{R}}_{0}$), defined as the average  number of susceptible individuals infected by a single infected individual during their infectious period  \cite{allen2008introduction}. 
These models can efficiently capture the optimal vaccination strategy ~\cite{wang2016statistical,heesterbeek2015modeling}, effective awareness program~\cite{eames2009networks,shams2014using,masuda2009immunization,miller2007effective}, efficient contact tracing~\cite{giordano2020modelling,aleta2020modelling} technique and suitable social distancing ~\cite{meidan2020alternating,vespignani2020modelling,weitz2020modeling} plan  to delay or eradicate the spread of the disease.
 \par
However, in absence of therapeutics, these types of strategies (social distancing, contact tracing etc.) might not annihilate the disease significantly if the infected or exposed individuals are not identified properly.  It is note worthy that achieving a significant reduction in disease incidence might not be at all possible unless the infected individuals are identified promptly as in most of disease a large portion of infected individuals shows mild symptoms or even no symptoms~\cite{who2016after, scarselli2020catastrophic, gandhi2020asymptomatic, furukawa2020evidence,heesterbeek2015modeling,dhillon2015community}. Also, a prolonged  and strict lock down may not be suitable for the  sustained and stable  economic condition of a country and an alternative solution by allowing the  partial opening of the business centers leading to enhanced public interaction is required for the economic stability. In this chaotic situation, the strategy like   aggressive testing  might  be quite efficient in identifying infected persons.}

\par Here, we propose an efficient test-kit  based control strategy (using high dimensional deterministic model) in  an infected population  to harness  the multifaceted cost of lock-down and social distancing. Under a systematic implementation of test-kits, we unveil, that it may significantly reduce the peak of infection as well as  the final outbreak size. The basic idea of our study lies within the  usage of  test-kits in which the number of production is assumed to be dependent on the current infection level  and by using the test-kit, the procedure of admitting the infected individual in the hospital is accelerated. A detailed investigation of the model enables us to accurately predict the relative reduction of the peak of the infection as well as final outbreak size in the presence of test-kit based strategy. 
\par A general consequence of such intervention strategy is relevant if someone considers the mobility of human population between the patches/communities connected with non-local heterogeneous networks. { In this backdrop, the key question we raise here, {\it what will be the ideal and optimal way of  distribution procedure of test-kits in meta communities?}
In the last century, a major development in the aviation as well as transportation networks expedites the epidemics of infectious disease. For instance, SARS and H1N1 Influenza originating from a local community spread across different countries within a few months ~\cite{colizza2007predictability,hufnagel2004forecast,fraser2009pandemic}. Another example,  COVID-19 originating from Wuhan, China on December, 2019  severely affected  different countries within very short periods (3-4 months). Therefore,  a suitable and optimal control of infected network is essential and important  for faster eradication of epidemic spread.       
Motivated by the relative importance of human mobility in spatial spread of disease, here we find the suitable way of distributing test-kits in complex connectivity of the communities in which the mobility of human will be determined by the diffusive migration \cite{belik2011natural,brockmann2013hidden,hens2019spatiotemporal,senapati2019impact,senapati2019cholera}. Note that, this meta-population network with migration is analogous to the reaction-diffusion dynamics \cite{colizza2007reaction,belik2011natural,colizza2008epidemic,Calvett2020Metapopulation} i.e., particles (here fraction of population) diffuse and interact. 
It is  well known that the suitable  intervention strategy  by partially controlling the network can reduce the prevalence of the entire system \cite{wang2016statistical,shams2014using, masuda2009immunization, miller2007effective}.   
For instance,  the infection spreads rapidly through the hubs \cite{hens2019spatiotemporal}. Therefore, vaccination in targeted  nodes can  dramatically resist the infection spreading ~\cite{madar2004immunization, chen2008finding,liu2016biologically,tanaka2014random} compared to the random immunization.    
In this background, we propose an optimal strategy (in higher diffusive strength) by means of the distribution of test-kits. Exploiting the microscopic information of a network, we are able to show that incorporating test-kits in high  degree patches (or patches having high betweenness centralities) from the onset of epidemics can effectively  reduce the peak of the infection, i.e., implication of test-kits in a certain fraction of  high degree patches (patches with high betweenness) will have almost the same impact (the percentage of  reduction in the peak of infection) compared to  the implication of  test-kits in each patch/community of the entire network. The proposed strategy acts like an effective immunization technique neglecting the information of the nodes which may have high prevalence in absence of test-kit implementations. To validate  our optimal strategy, we  consider two real networks: the global connectivity pattern of flights and the transportation networks of the city Kolkata.} 
\section{Single-node Model Description}
\label{sec:single_model}
\par We consider a generic S-E-I-R type (Susceptible-Exposed-Infected-Recovered) model in deterministic set up and extend it by introducing two state variables: one is $H$, denoting number of hospitalized persons and another is $K$, which represents the number of test-kit. The human population is categorized into five compartments: susceptible ($S$), exposed ($E$), infected ($I$), hospitalized ($H$) and recovered ($R$) depending on the current health status of the individuals. The individuals who are susceptible to a disease become exposed if they experience close contacts with the infected individuals. The force of infection is given by 
$\beta\frac{I}{N}$, 
where $\beta$ is the rate of disease transmission from infected to susceptible and $N$ is number of the total human population. The individuals in the exposed compartment do not have the ability to transmit the disease among other susceptible individuals as the pathogen abundance in their body is not sufficient for active transmission. Sometimes this compartment is therefore called latent. At the end of the latent period 
(i.e. $\frac{1}{\sigma}$), 
the exposed individuals move to the infected compartment. In this stage, individuals show the symptoms of the disease and are capable of spreading the infection. The infected individuals then move to the hospitalized compartment at a rate 
$\alpha(K)$
 and then get recovered from the disease at a rate 
 $\gamma$. 
 We consider the disease to be non-fatal and therefore neglect the disease induced mortality. It is assumed that the process of hospitalization depends on the availability of the test-kit. With the help of the test-kit, the undetected infected individuals are tested and among them who are tested positive, admitted to the hospital. The implementation of testing procedure wit the help of test-kit basically speeds up the hospitalization process. As a result, the infected individuals get lesser time to transmit the disease to the susceptible population. For the sake of simplicity, we consider the rate of hospitalization as a linear function of the available test-kit, i.e.,
$\alpha(K)=\alpha_{0}+\alpha_{1}K$, where $\alpha_{0}$ 
denotes the rate of hospitalization in the absence of any test-kit, 
$\alpha_{1}$ 
is the effectiveness of the test-kit. We assume that the test-kit is produced as  proportion
($\xi$) 
of the number of current infected individuals and looses its efficacy at a rate $\chi$. Note that, since we consider a disease outbreak situation and the duration of outbreak is generally shorter in compared to the human demographic process, therefore we neglect the demographics in our study.
At any instant of time $t$, the rate of change of the number of individuals in the above-mentioned human compartments and that of the number of test-kit can be expressed mathematically as a set of ordinary differential equations as follows:       
\begin{eqnarray}
\label{eq:single_model_S} 
\frac{dS}{dt} &=& \displaystyle -\beta\frac{S I }{N},\\
\frac{dE}{dt} &=& \displaystyle \beta\frac{S I }{N} - \sigma E \label{eq:single_model_E},\\
\frac{dI}{dt} &=& \displaystyle  \sigma E - \alpha(K) I, \\   
\frac{dH}{dt} &=&  \alpha(K) I -\gamma H, \label{eq:single_model_H} \\ 
\label{eq:single_model_R} 
\frac{dR}{dt} &=&  \gamma H, \\   
\frac{dK}{dt} &=&  \xi I - \chi K.
\label{eq:single_model_K}   
\end{eqnarray}

\subsection{Basic Reproduction Number}
\label{sec:basic_reproduction_number}
Basic reproduction number is a key quantity in epidemiology which measures the severity of the disease. It basically indicates, for a disease, on average how many new cases is being generated from a typical infected individual during his/her infectious period~\cite{van2002reproduction,diekmann2010construction}. We follow the next-generation matrix approach~\cite{van2002reproduction,diekmann2010construction} to calculate the basic reproduction number ($\mathcal{R}_{0}$) 
for our model Eqns.\ \eqref{eq:single_model_S}-\eqref{eq:single_model_K}.
\par Following this approach, we first rearrange our model (Eqns.\ \eqref{eq:single_model_S}-\eqref{eq:single_model_K}) so that the infected compartments appear first in the set of equations. Now our system can be written in the vector-form as:
\begin{equation*}
\frac{d\textbf{x}}{dt}=\mathcal{F}-\mathcal{V},
\end{equation*} where, $\textbf{x}=[E,I,S,H,R,K ]^{T}$ and $\mathcal{F}$ and $\mathcal{V}$ are given by,
\[
\mathcal{F}=
\left[{\begin{array}{cc}
	\beta \frac{S I}{N}	\\
	0\\
	0\\
	0\\
	0\\
	0\\
	\end{array}} \right],
{\rm and}~~\mathcal{V}=
\left[{\begin{array}{cc}
	\sigma E\\
	-\sigma E + \alpha(K)I\\
	\beta\frac{SI}{N}\\
	-\alpha(K)I+\gamma H\\
	-\gamma H\\
	-\xi I +\chi K\\
	\end{array}} \right].
\]

Now the partial derivatives of $\mathcal{F}$ and $\mathcal{V}$ are evaluated with respect to the infected compartments $E$ and $I$ at disease-free equilibrium ($\mathcal{E}_{0}=(N,0,0,0,0,0)$) to obtain the new infection matrix $F$ and transmission matrix $V$ as follows: 
\[
F=
\left[{\begin{array}{cc}
	\frac{\partial }{\partial E}(\frac{\beta SI}{N} ) & \frac{\partial}{\partial I} (\frac{\beta SI}{N} )\\
	\frac{\partial }{\partial E}(0) & \frac{\partial }{\partial I}(0)\\
	\end{array}} \right]_{\rm{\mathcal{E}_{0}}}
=
\left[{\begin{array}{cc}
	0 & \beta\\
	0  & 0 \\
	\end{array}} \right],
\]
and 

\[
V=
\left[{\begin{array}{cc}
	\frac{\partial }{\partial E}(\sigma E) & \frac{\partial}{\partial I} (\sigma E)\\
	\frac{\partial }{\partial E}[-\sigma E+(\alpha_{0}+\alpha_{1}K)I] & \frac{\partial }{\partial I}[-\sigma E+(\alpha_{0}+\alpha_{1}K)I]\\
	\end{array}} \right]_{\rm{\mathcal{E}_{0}}}
=
\left[{\begin{array}{cc}
	\sigma  & 0\\
	-\sigma  & \alpha_{0}\\
	\end{array}} \right].
\] 

The basic reproduction number is defined as the spectral radius of the matrix $FV^{-1}$, i.e., $\mathcal{R}_{0}=\rho(FV^{-1})$, where $\rho(A)$ denotes the spectral radius of the matrix $A$. In our case, the basic reproduction number ($\mathcal{R}_{0}$) is given by,
\begin{equation}
\mathcal{R}_{0}=\frac{\beta}{\alpha_{0}}.
\label{eq:basic_reproduction}
\end{equation}
It is to be noted that $\mathcal{R}_{0}$ is used as a threshold quantity for the stability of disease-free equilibrium. It can be proved easily (using linear stability analysis) that if $\mathcal{R}_{0}<1$, then the disease-free equilibrium is locally asymptotically stable and unstable whenever $\mathcal{R}_{0}>1$. For our case, the disease-free equilibrium become unstable or in other words, the disease outbreak starts if $\beta>\alpha_{0}$.
\begin{figure*}[ht]
	\includegraphics[width= 1\textwidth]{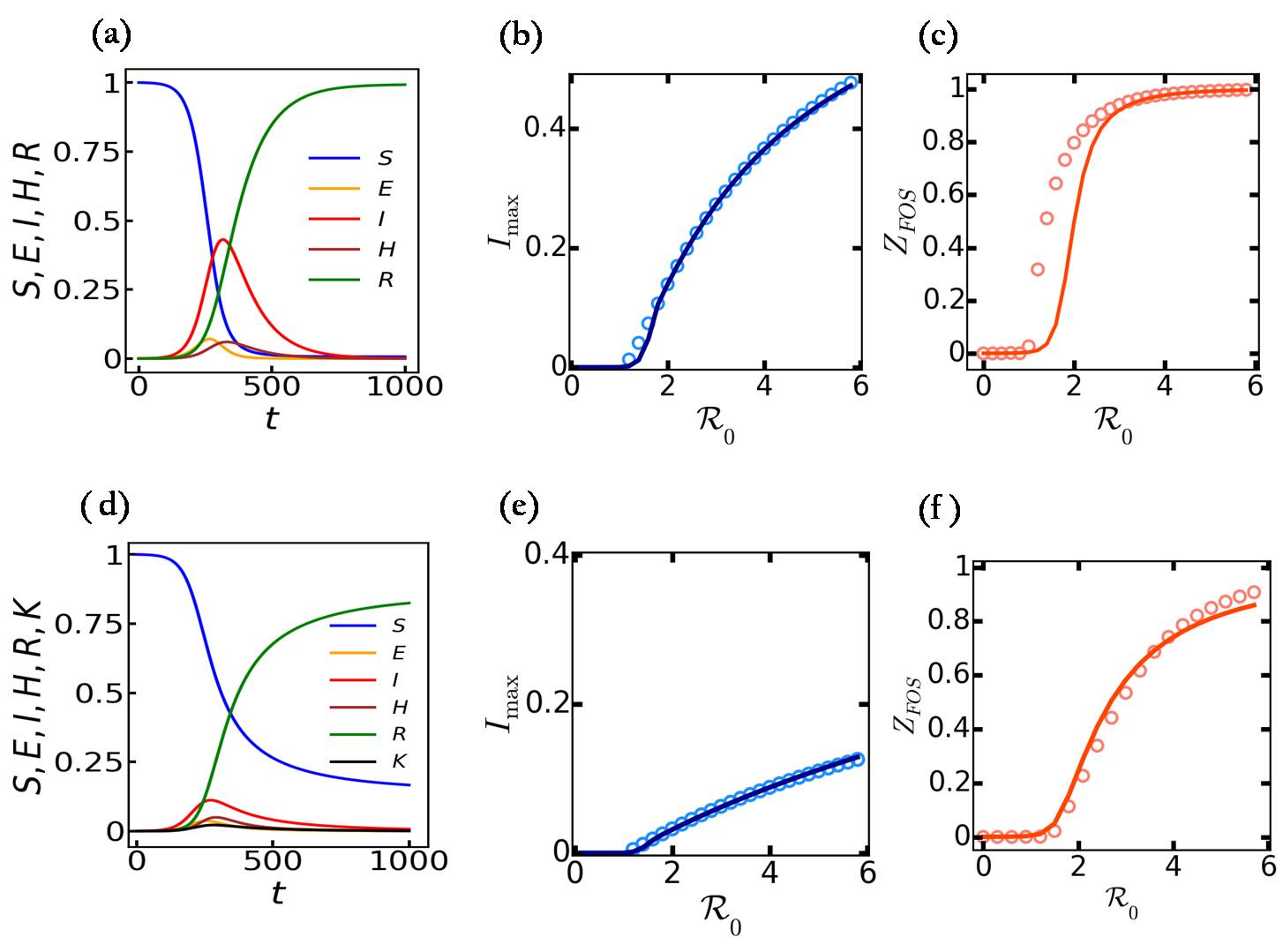}
	\caption{{\bf Dynamics of the single-node model  in the absence and in the presence of test-kit.} (a) Time evolution of all the compartments in the absence of test-kit (i.e. $\alpha_{1}=0$).
		The basic reproduction number, $\mathcal{R}_{0}$ in this case is $3$. (b) Peak of the infection ($I_{\rm max}$) is plotted by varying $\mathcal{R}_{0}$. The solid blue line represents the model outcomes and the blue circles are the values of $I_{\rm max}$ obtained from the analytical expression~\eqref{eq:I_max_without_kit}. (c) Final outbreak size ($Z_{\rm FOS}$) is plotted against $\mathcal{R}_{0}$ for $\alpha_1=0$. The solid red line is the model output and the red circles are the values of $Z_{\rm FOS}$ obtained from solving the transcendental equation~\eqref{eq:Zfos_final}. (d) Time evolution of all the human compartments and also of the test-kit for $\mathcal{R}_{0}=3$. 
		(e) $\mathcal{R}_{0}$ vs. $I_{\rm max}$ for $\alpha_1\neq0$. The solid blue line represents the model outcome and the blue circles are obtained from~\eqref{eq:I_max_with_kit}.  (f) $\mathcal{R}_{0}$ vs. $Z_{\rm FOS}$ for $\alpha_1\neq0$. The solid red line is the model output and the red circles are the values of $Z_{\rm FOS}$ obtained by  transcendental equation~\eqref{eq:Zfos_with_final}.   The parameter values are taken as: $\sigma=0.1$, $\alpha_{0}=0.01$. $\alpha_{1}=0.0001$, $\gamma=0.07$, $\xi=0.02$, $\chi=0.1$. The parameter $\beta$ is varied in the range $[0,0.06].$}
	\label{single_model_plot}
\end{figure*}
\subsection{Peak of Infection}
\label{sec:peak_of_infection}
In outbreak situation (i.e. for $\mathcal{R}_{0}>1$), the number of infected population ($I(t)$) initially increases and reaches its maximum and in the subsequent time it starts to decrease and ultimately goes to zero at the end of the outbreak. Here we derive the analytical expression for the maximum of infected population ($I_{\rm max}$) as a function of basic reproduction number, $\mathcal{R}_{0}$.
\par Let us denote $X=H+R$. Adding the equations  \eqref{eq:single_model_H} and \eqref{eq:single_model_R}, we have,
\begin{eqnarray}
\frac{dX}{dt}&=&(\alpha_{0}+\alpha_{1}K)I.
\label{eq:I_max_1} 
\end{eqnarray}
\par \textbf{Case 1:} { At first we consider the case where test-kits are not introduced i.e. $\alpha_{1}=0$}.
We divide the equation of susceptible ($S$)  (Eqn.\ \eqref{eq:single_model_S}) by Eqn.\ \eqref{eq:I_max_1},
\begin{eqnarray}
\frac{dS}{dX}&=&-\frac{\beta S}{\alpha_{0}N}\nonumber.
\end{eqnarray}
Solving the above equation we get explicit solution of $S(t)$ as, 
\begin{eqnarray}
S(t)=S(0)e^{-\frac{\beta X}{\alpha_{0}N}} \label{eq:S_X}.
\end{eqnarray} 
Now we express the variables $I$ and $E$ in terms of $X$. {Plugging the first order and second order derivatives of $X$ into    Eqn.\ \eqref{eq:single_model_E}  we may write}
\begin{eqnarray}
E &=&\frac{1}{\sigma} \bigg(\frac{1}{\alpha_{0}}\frac{d^2X}{dt^2}+\frac{dX}{dt}\bigg).
\label{eq:E_X}
\end{eqnarray}
Substituting the expressions of $S,~E,~I$ from the Equations \eqref{eq:S_X} - \eqref{eq:E_X} and Eqn.\ \eqref{eq:I_max_1}  into the constraint relation $S+E+I+X=N$, we obtain \cite{das2020covid}
\begin{equation*}
S(0)e^{-\frac{\beta X}{\alpha_{0}N}}+\frac{1}{\sigma} \bigg(\frac{1}{\alpha_{0}}\frac{d^2X}{dt^2}+\frac{dX}{dt}\bigg)+\frac{1}{\alpha_{0}}\frac{dX}{dt}+X-N=0.
\end{equation*}
Simplifying the above equation we get a second order differential equation of $X$ as follows:
\begin{equation*}
\frac{1}{\sigma\alpha_{0}}\frac{d^2X}{dt^2}+ \bigg(\frac{1}{\sigma}+\frac{1}{\alpha_{0}}\bigg)\frac{dX}{dt}+(X+S(0)e^{-\frac{\beta X}{\alpha_{0} N}}-N)=0.  
\end{equation*}
Let us denote $x=\frac{X}{N}$, and $y=\frac{dx}{dt}.$
By this transformation the above second order differential equation can be reduced to a system of first order differential equations, 
\begin{eqnarray}
\begin{array}{llll}
\displaystyle \frac{dx}{dt}&=&\displaystyle y,\\
\displaystyle \frac{dy}{dt}&=& \displaystyle -(\alpha_{0}+\sigma)y-\alpha_{0}\sigma(x+e^{-\mathcal{R}_{0}x}-1).
\end{array}
\label{eq:I_max_2} 
\end{eqnarray} 
Assuming,  that $E$ and $I$ reach the peaks almost at the same time ($t=\tau$) we can write 
$\frac{dE}{dt}=\frac{dI}{dt}=0$. 
Therefore the time derivative of the constrain function $\frac{dS}{dt}+\frac{dE}{dt}+\frac{dI}{dt}+\frac{dX}{dt}=0$ reduces to $\frac{dS}{dt}+\frac{dX}{dt}=0$ at $t=\tau$.  
This modified relation leads us into
$S(\tau)=\frac{N}{\mathcal{R}_{0}}$.
Also from the Eqn.\  \eqref{eq:S_X} 
we can express  
$X(\tau)=\frac{N}{\mathcal{R}_{0}}\ln \mathcal{R}_{0}$.

Since $\frac{dy}{dt}=\frac{\alpha_{0}}{N}\frac{dI}{dt}$, the maximum value of $I$ can be obtained indirectly by  equating $\frac{dy}{dt}$ to $0$ from Eqn.\ \eqref{eq:I_max_2}. We also assumed that $I$ reaches its peak at $t=\tau$, which implies that $y$ also reaches its maximum at that point. Therefore, the maximum value of $y$ which is obtained at the point $t=\tau$ (by $\frac{dy}{dt}=0$ at $t=\tau$ from Eqn.\ \eqref{eq:I_max_2}) is given by,
\begin{equation*}
y(\tau)=-\frac{\sigma \alpha_{0}}{\sigma +\alpha_{0}}\bigg(x(\tau)+e^{-\mathcal{R}_{0}x(\tau)}-1\bigg).
\end{equation*}  
Now using $y(\tau)=\frac{\alpha_{0}I(\tau)}{N}$ and $X(\tau)=\frac{N}{\mathcal{R}_{0}}\ln \mathcal{R}_{0}$ in the above equation we finally get the expression for $I_{\rm max}~(=I(\tau))$ as follows:
\begin{equation}
I_{\rm max}=\frac{\sigma N}{\sigma+\alpha_{0}}\bigg(1-\frac{1+\ln \mathcal{R}_{0}}{\mathcal{R}_{0}}\bigg).
\label{eq:I_max_without_kit}
\end{equation}
\par \textbf{Case 2: $\alpha_{1}\neq0$}, i.e. in the presence of test-kit.
\par In this case, since $K$ is changing over the time, it is very difficult to obtain exact expression for $I_{\rm max}$. Therefore, for the sake of simplicity we assume that $K(t)=K^*$  for all $t>0$. 
Following the similar procedure as described for \textbf{Case 1}, we obtain $I_{\rm max}$ as,
\begin{equation*}
I_{\rm max}=\frac{\sigma N}{\sigma+\alpha_{0}+\alpha_{1}K^{*}}\bigg(1-\frac{1+\ln \frac{\mathcal{R}_{0}}{1+\frac{\alpha_{1}}{\alpha_{0}}K^*}}{\frac{\mathcal{R}_{0}}{1+\frac{\alpha_{1}}{\alpha_{0}}K^{*}}}\bigg).
\end{equation*}        
Now we propose suitable value for $K^*$ as $K^*=aI_{\rm max}$, where $a$ is a non-negative constant. That means we assume that $K^*$ can be obtained by multiplying $I_{\rm max}$ by a suitable constant $a$. Now the above expression for $I_{\rm max}$ becomes 
\begin{equation}
I_{\rm max}=\frac{\sigma N}{\sigma+\alpha_{0}+\alpha_{1}aI_{\rm max}}\bigg(1-\frac{1+\ln \frac{\mathcal{R}_{0}}{1+\frac{\alpha_{1}}{\alpha_{0}}aI_{\rm max}}}{\frac{\mathcal{R}_{0}}{1+\frac{\alpha_{1}}{\alpha_{0}}aI_{\rm max}}}\bigg).
\label{eq:I_max_with_kit}
\end{equation}     
It is to be noted that the Eqn.\ \eqref{eq:I_max_with_kit} is a transcendental equation of $I_{\rm max}$. 

\subsubsection*{Numerical Results of  $I_{\rm max}$ and analytical validation}
In absence of test-kits, the time evolution of all the normalized  variables are shown in the Fig.\ \ref{single_model_plot}(a). The infection rate $\beta$ is fixed at $0.03$ i.e.  
$\mathcal R_0 =3$ 
(see Eqn.\ (\ref{eq:basic_reproduction}) for analytical proof). The rest of the parameters are mentioned in Fig.\ \ref{single_model_plot}. In this parameter set-up the infection ($I$, red line) 
gets a maximum value around 
$0.4$ 
which occurs  at 
$\sim210$ 
days from the onset of infection. 
In presence of test-kits,  exposed individuals as well as infected individuals are significantly dropped 
at ($I_{\rm max}\sim 0.15$) 
shown in Fig.\ \ref{single_model_plot}(d) and  
$20\%$ 
susceptible population (blue) remain unperturbed. As a result the recovered individuals 
($R$) 
saturates around the 
$0.8$ (green).
We have also checked the maximum of the infection 
($I_{\rm max}$) 
as a function of  basic reproduction number 
$\mathcal{R}_{0}$ 
in absence of test-kits  (Fig.\ \ref{single_model_plot}(b)).  Here  analytically calculated 
$I_{\rm max}$  
(Eqn.\ (\ref{eq:I_max_without_kit})) is almost matched (blue circles) with the numerical result(solid blue line). Interestingly, in presence of test-kits $I_{\rm max}$ is significantly reduced  shown in  Fig. \ref{single_model_plot}(e)  with solid blue lines. The analytically obtained transcendental equation (\ref{eq:I_max_with_kit})  validates  the numerical results extracted from the solution of   the coupled differential equations using RK4 routine. Here the transcendental equation is solved  for a wide range of $a$ ($a\in[0.1,0.2]$), 
the average result is shown with blue circle. 
\subsection{Final Outbreak Size}
\label{sec:outbreak_size} 
When an outbreak starts, the susceptible population tend to decrease over the time. However, the chain of disease transmission is interrupted due to the reduction in infected population. It is to be noted that there is always a portion of susceptible population who are able to avoid the infection~\cite{keeling2008modeling}. Final outbreak size ($Z_{\rm FOS}$) quantifies the portion of total population got infected at the end of an outbreak of disease. Speaking in term of susceptible population, it gives the portions of susceptibles who are able to protect themselves from the infection in a particular disease outbreak. $Z_{\rm FOS}$ basically provides the asymptotic behaviour of the system. In this section, we present the analytical treatment of $Z_{\rm FOS}$.     
\par We assume that initially all the population is in susceptible state, i.e., $S_{0}=N$. From  the model equations  \eqref{eq:single_model_S}- \eqref{eq:single_model_E} we see that,
\begin{eqnarray}
\nonumber    
\frac{dS}{dt} + \frac{dE}{dt}&=&-\sigma E < 0.
\label{OS_1}      
\end{eqnarray}
This implies that $S(t)+E(t)$ is decreasing function of $t$. Since $\sigma>0$, and also $S(t)+E(t)$ is non-negative function, $E(t)$ should tend to zero as $t \to \infty$, i.e., $E_{\infty}=0.$ Again we see that,
\begin{eqnarray}
\frac{dS}{dt} +\frac{dE}{dt}+ \frac{dI}{dt}&=&-(\alpha_{0}+\alpha_{1}K)I \leq 0.
\label{eq:Zfos_1}      
\end{eqnarray}
Using the similar argument as above, we have $I \to 0$, as $t \to \infty$, i,e $I_{\infty}=0$.
\par From equation Eqn.\ \eqref{eq:Zfos_1}, we can write,
\begin{eqnarray}
\nonumber    
-\int_{0}^{\infty}(\alpha_{0}+\alpha_{1}K)I dt&=& \int_{0}^{\infty} d(S+E+I).
\label{OS_3}      
\end{eqnarray}
On simplifying the above equation and using $E_{\infty}=0$, $I_{\infty}=0$ we finally get,
\begin{eqnarray}
\alpha_{0} \bar{I} + \alpha_{1}\int_{0}^{\infty}KIdt &=& (S_{0}-S_{\infty})+E_{0}+I_{0},
\label{eq:Zfos_2}
\end{eqnarray}
where $\bar{I}$ is defined as $\int_{0}^{\infty}I(t) dt$.

\textbf{Case 1:} $\alpha_{1}=0$, i.e., in the absence of any test-kit.

Integrating \eqref{eq:single_model_S}, we have
\begin{eqnarray}  
\nonumber  
\int_{0}^{\infty} \frac{\dot{S}}{S} dt  &=& - \frac{\beta}{N} \int_{0}^{\infty} I dt.
\end{eqnarray}
Using the value of $\bar{I}$ from Eqn.\ \eqref{eq:Zfos_2}, 
\begin{eqnarray}
\nonumber
\ln(\frac{S_{\infty}}{S_{0}}) 
&=& - \frac{\mathcal{R}_{0}}{N} \bigg(S_{0}-S_{\infty}+E_{0}+I_{0}\bigg)
\end{eqnarray}
Let us denote $s_{\infty}=\frac{S_{\infty}}{S_{0}}$. Then from the above equation the we obtain the equation of $s_{\infty}$ as follows:
\begin{eqnarray}
\nonumber
s_{\infty} &=&  e^{- \mathcal{R}_{0} \left[(1- s_{\infty})+\frac{(E_{0}+I_{0})}{S_{0}}\right]}.
\end{eqnarray}
Now we define final outbreak size ($Z_{FOS}$) as $ Z_{\rm FOS}=1-s_{\infty}$. Finally the expression of $Z_{\rm FOS}$ is given by,
\begin{eqnarray}    
Z_{\rm FOS} &=& 1- e^{- \mathcal{R}_{0} \left[Z_{\rm FOS}+\frac{(E_{0}+I_{0})}{S_{0}}\right]}.
\label{eq:Zfos_final}      
\end{eqnarray}
\par \textbf{Case 2:} $\alpha_{1} \neq 0$, i.e., in the presence of test-kit.

We recall the equation Eqn.\ \eqref{eq:Zfos_2},
\begin{eqnarray}
\nonumber    
\alpha_{0} \bar{I} + \alpha_{1}\int_{0}^{\infty}KI  dt &=& (S_{0}-S_{\infty})+E_{0}+I_{0}. 
\label{OS_8}      
\end{eqnarray}
We now evaluate the upper bound of the integral $\int_{0}^{\infty}K(t)I(t)dt$. 
Note that $I\leq N$ is true from the model for all $t>0$, we have, 
\begin{eqnarray} 
\int_{0}^{\infty}K(t)I(t)  dt \leq N \int_{0}^{\infty}K  dt= N \bar K.
\label{eq:Zfos_with_1}    
\end{eqnarray}
Now we calculate $\bar K $ from the Eqn.\ \eqref{eq:single_model_K} 
\begin{eqnarray}
\nonumber    
\int_{0}^{\infty}\frac{dK}{dt} dt &=& \xi \int_{0}^{\infty} I dt - \chi \int_{0}^{\infty} K dt.
\label{OS_10}      
\end{eqnarray}
From the above equation, we obtain $\bar{K}$ as follows:
\begin{eqnarray}
\nonumber
\bar K = \frac{1}{\chi} \bigg(\xi \bar I  + K_{0}-K_{\infty}\bigg).
\label{eq:Zfos_with_2}
\end{eqnarray}
Plugging the above expression of $\bar{K}$ in Eqn.\ \eqref{eq:Zfos_with_1}, we get the upper bound of the integral, 
\begin{eqnarray}    
\int_{0}^{\infty} K(t)I(t) dt \leqslant \frac{N}{\chi} \bigg(\xi \bar I  + K_{0}-K_{\infty} \bigg).
\label{eq:Zfos_with_3}     
\end{eqnarray}
Now we use Eqn.\ \eqref{eq:Zfos_with_3} in the Eqn.\ \eqref{eq:Zfos_with_1} and get the following inequality,
\begin{eqnarray}   
\bar I \geqslant \frac{1}{(\alpha_{0}+\alpha_{1}\frac{N \xi}{\chi})}\left[ (S_{0}-S_{\infty})+(E_{0}+I_{0})+\alpha_{1}\frac{N}{\chi}( K_{\infty}-K_{0})\right].
\label{eq:Zfos_with_4}     
\end{eqnarray}
Following the same procedure as described in \textbf{Case 1}, here we also define $s_{\infty}=\frac{S_{\infty}}{S_{0}}$. 
Then the above inequality can now be expressed in terms of $s_{\infty}$ as, 
\begin{eqnarray} 
\nonumber   
s_{\infty} \leqslant  e^{-\mathcal{R}_{0} \frac{\alpha_{0}}{(\alpha_{0}+\alpha_{1}\frac{N \xi}{\chi})} \left[(1-s_{\infty})+\frac{(E_{0}+I_{0})}{N} +\frac{\alpha_{1}}{\chi}( K_{\infty}-K_{0}) \right]}.
\label{OS_13}      
\end{eqnarray}
Finally the we define final outbreak size
 ($Z_{\rm FOS}$) as $Z_{\rm FOS}=1-s_{\infty}$ 
 and obtain the following transcendental equation of $Z_{\rm FOS}$:   
\begin{eqnarray}    
Z_{\rm FOS} \geqslant 1- e^{- \mathcal{R}_{0} \frac{\alpha_{0}}{(\alpha_{0}+\alpha_{1}\frac{N \xi}{\chi})} \left[Z_{\rm FOS}+\frac{(E_{0}+I_{0})}{N} +\frac{\alpha_{1}}{\chi}( K_{\infty}-K_{0}) \right]}.  
\label{eq:Zfos_with_final}      
\end{eqnarray}
\subsubsection*{Numerical Results  of $Z_{\rm FOS}$ and analytical validation}
In absence of test-kits ($\alpha_1=0$), the final outbreak size ($Z_{\rm FOS}$) 
is plotted against $\mathcal{R}_{0}$ 
shown in the Fig.\ \ref{single_model_plot}(c). 
The solid red line is the numerically integrated data of the model equations and the red circles are the values of $Z_{\rm FOS}$ 
obtained from the transcendental equation (\eqref{eq:Zfos_final}, $E_0<<N, ~ I_0<<N$). For  
$\alpha_1\neq0$ 
i.e., in presence of test-kits, the final outbreak size is shown in Fig.\  \ref{single_model_plot}(f) with solid red line and the semi-analytically obtained  
$Z_{\rm FOS}$ (see transcendental equation \eqref{eq:Zfos_with_final}, where $K_{\infty}=K_0=0$) is shown with red circles. The transcendental equation is solved by using a wide range of $N$:  $N\in[300,1000]$.
 Clearly,  in lower $\mathcal{R}_{0}$ ($\sim 2-4$), a large fraction of susceptible population can avoid the infection due to the rigorous-testing procedure.  Note that, the  time series (Fig.\ \ref{single_model_plot} a,d) of the compartmental variables are plotted at $\mathcal{R}_{0}=3$.


\section{Data Fitting: A case study on COVID-19}
To validate our proposed model (\eqref{eq:single_model_S}-\eqref{eq:single_model_K}), we choose the cumulative daily COVID-19 hospitalized data for three sates of United States of America (U.S.A), namely Maryland, Ohio, and New york. The data is collected from the website (https://covidtracking.com/data/download). For Maryland and Ohio, we use the data during the period $5^{th}$ March, 2020 to $5^{th}$ September, 2020 (i.e. 185 data points) and for New York, we use the same during the period $4^{th}$ March, 2020 to $5^{th}$ September, 2020 (i.e. 186 data points). The total population in these three states are obtained from (https://www2.census.gov).
\par We estimate four unknown model parameters: (i) the disease transmission rate ($\beta$), (ii) effectiveness of test-kit ($\alpha_{1}$), (iii) rate of production of kit ($\xi$), (iv) rate of losing efficacy of test-kit ($\chi$) by fitting our model to cumulative daily COVID-19 hospitalized data for three states of U.S.A.
\par At any time instant $t$, the cumulative number of hospitalized persons from the model is given by,
\begin{eqnarray}
C(t,\Theta)=C(1)+\int_{1}^{t} \bigg(\alpha_{0}+\alpha_{1}K(\tau)\bigg)I(\tau)d\tau,
\end{eqnarray}
where $\Theta=\{\beta, \alpha_{1}, \xi, \chi\}$ and $C(1)$ denotes the cumulative number of hospitalized persons at first day (i.e. at $t=1$).
\par
\begin{figure*}[h]
	{\includegraphics[width= 1\textwidth]{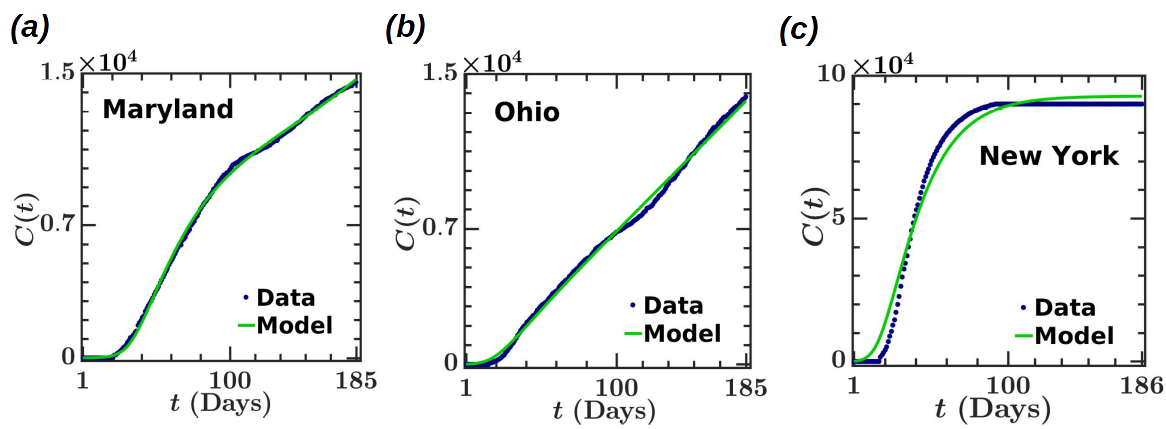}}
	\caption{{\bf Model fitting with real data.} (a)-(c) Output of the model fitted with cumulative number of daily COVID-19 hospitalized persons in Maryland, Ohio, and New York respectively. In figures (a) and (b), the starting time point, i.e., $t=1$ represents the date $5^{th}$ March, 2020 and in figures (c), $t=1$ represents the date $4^{th}$ March, 2020. In all the figures (a)-(c), the dates representing the end points ($t=185$ for (a), (b) and $t=186$ for (c)) are same, i.e., $5^{th}$ September, 2020. The blue dot represents the discrete data point and green curve represents the model solution in all the figures. The fixed parameters are taken same as in Fig.\ \ref{single_model_plot} except $\alpha_{0}=0.1$.}
	\label{Fig:fitting}
\end{figure*}

 We perform our model fitting by using in-built function \textit{lsqnonlin} in MATLAB (Mathworks, R2014a) to minimize the sum of square function. In our case, the sum of square function $SS(\Theta)$ is given by,

\begin{eqnarray}
SS(\Theta)=\displaystyle \sum_{i =1}^{n} \big(C^{d}(t_{i})-C(t_{i},\Theta)\big)^2,
\end{eqnarray}
where, $C^{d}(t_{i})$ is the actual data at $t_{i}^{th}$ day and $n$ is the number of data points. The model fitting to the cumulative daily hospitalized data for the three states is displayed in Fig.\ \ref{Fig:fitting}. {The blue dots in each figure capture the cumulative number of hospitalized data where as the green line is the corresponding model output. The values of the \textit{best-fit} parameters ($\hat{\beta},\hat{\alpha_{1}}, \hat{\xi}, \hat{\chi}$) are given in  supplementary material (SM) Sec.\ \ref{Best-fit parameters}, Table 1. From Fig.\ \ref{Fig:fitting}, it is clear that our model captures the real outbreak scenario quite well. }

\section{Optimal Intervention strategy in heterogeneous  networks}
\label{Network Model}
Now we extend our  approach to a metapopulation network \cite{belik2011natural,brockmann2013hidden,hens2019spatiotemporal}. We consider  a heterogeneous network of $M$ patches, i.e., $M$ number of communities/nodes. Initially we assume,   there will be   $N_n (n=1,2,...,M)$ number of susceptible people in each community. We are also assuming, a small fraction of certain communities is infected. Note that,  total population,
$\mathcal{N_{\rm tot}}=\sum_{n=1}^{M} N_n$
is conserved. Considering the dispersion through diffusion of susceptible ($S_n$), exposed ($E_n$), infected ($I_n$)
and recovered ($R_n$)
individuals, we may write the coupled network equations as
\begin{eqnarray}
\begin{array}{llll}
\frac{dS_{n}}{dt} &=& \displaystyle -\beta_{n} \left (\frac {S_{n} I_{n}}{N_{n}}\right)   + \frac {\epsilon}{d_{n}} \sum_{m=1}^{\mathrm{M}}  A_{nm} (S_{m}-S_{n}), \\
\frac{dE_{n}}{dt} &=& \displaystyle \beta_{n} \left (\frac {S_{n} I_{n}}{N_{n}}\right) - \sigma E_{n} + \frac {\epsilon}{d_{n}} \sum_{m=1}^{\mathrm{M}}  A_{nm} (E_{m}-E_{n}),\\
\frac {dI_{n}}{dt} &=& \displaystyle \sigma E_{n} -({\alpha_0+ g(K)}) I_{n} + \frac {\epsilon}{d_{n}} \sum_{m=1}^{\mathrm{M}}  A_{nm} (I_{m}-I_{n}), \\
\frac {dH_{n}}{dt}&=& \displaystyle ({\alpha_0+ g(K)}) I_{n} - \gamma H_{n}, \\
\frac {dR_{n}}{dt} &=& \displaystyle \gamma H_{n} +\frac {\epsilon}{d_{n}} \sum_{m=1}^{\mathrm{M}} A_{nm} (R_{m}-R_{n}),  \\
\frac {dK}{dt} &=& \displaystyle \xi \sum_{n=1}^{\mathrm{M}} I_{n} - \chi K.   
\end{array}
\label{eq:network_model}   
\end{eqnarray}
Here, $ Anm$ is the element of the  {weighted  adjacency matrix $A$ revealing the connectivity pattern among the patches}. 
$d_n=\sum_{m=1}^M A_{nm}$  
is the degree (number of neighbours) of the 
$n^{\rm th}$ patch. 
The migration is designed across population diffusion from one patch to another through the diffusive term 
$\sum_{m=1}^{\mathrm{M}}  A_{nm} (X_{m}-X_{n})$ 
connected through four compartments 
${X: S_n,E_n,I_n,R_n}$. 
$\epsilon$ determines the strength of the migration and $d_n$ in the denominator determines the average mean-field of the four compartments described above. 
We have produced the number of  kits as a proportional to the total  infected individuals 
($\sum_{n=1}^{\mathrm{M}} I_{n}$) 
in which a fraction of kits will be used in an infected compartment  by the function 
$g(K)=\alpha_{1}\times p_n K$. 
If the test-kits are equally distributed in each patch, the term $p_n$ will be equal to the inverse of the size of the network, i.e., $p_n=\frac{1}{M}$.  
If  the test-kits are distributed in $l$ patches according to their connectivity pattern i.e.\ the degree, one can write $g(K)=\alpha_{1}\times p_n K = \alpha_{1}\times \frac{d_n}{\sum_{n=1}^{l}{d_n}} K$. 
Hiring, these features into the compartment(s), we seek an  efficient distribution of kits within the communities for the optimal reduction of prevalence and peak of infection of the entire connected patches. Exploiting the micro structures of the network we identify  the key and influential patches in the  network where the test-kits will be provided to decrease the peak of infection largely. In the next sections, we will show, such fractional but potential intervention strategy may indeed reduce the peak of infection significantly and it can  also decrease the final outbreak size. We expect (under certain conditions), the test-kits distributed in  fraction of specific patches  will have same impact if we equally distribute the test-kits in each node. To quantify the total normalized infection (summed over  all the patches) we use $\mathcal{I}=\frac{\sum_{n=1}^{M} I_n}{\mathcal{N_{\rm tot}}}$. 
In a similar way,  we define the  normalized  final  outbreak size   as 
$\mathcal{Z}_{\rm FOS}=\frac{\sum_{n=1}^{M} {R}_n}{\mathcal{N_{\rm tot}}}$ 
i.e. the ratio of total recovered individual with respect to the total population. 

\begin{figure}
	\includegraphics[width=1\textwidth]{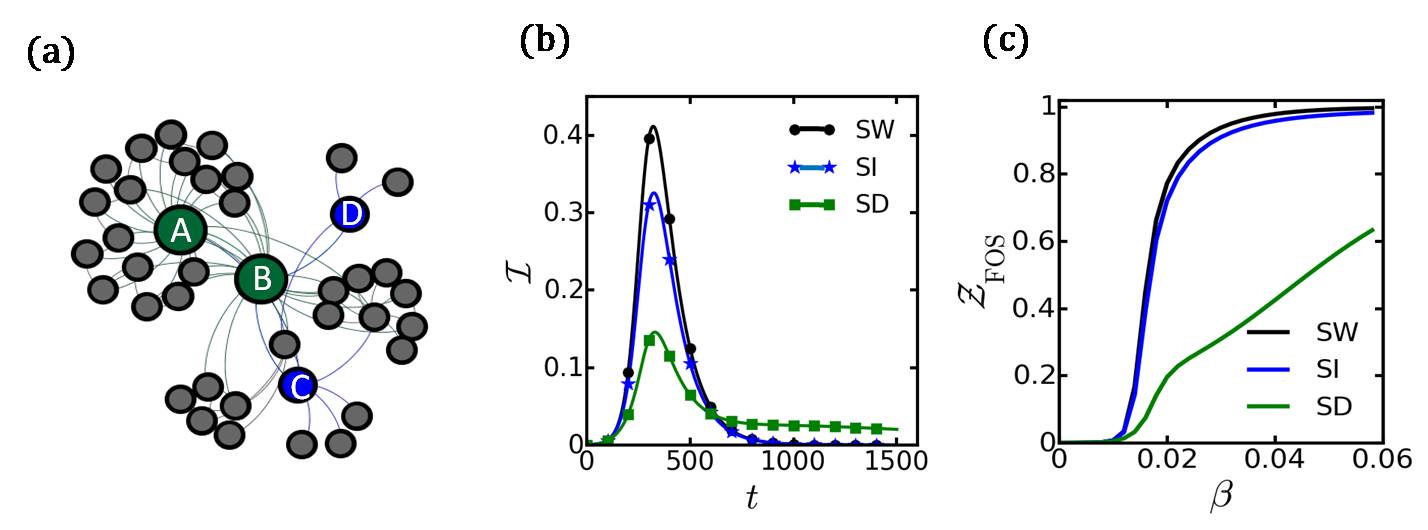}
	\caption{{\bf Network structure and optimal testing kit based control strategy.} (a) A Small network of $40$ nodes is used. Two strategies are used : either patches A and B (green) where kits are degree-wise distributed (SD) or C and D (blue) where test-kits are equally distributed (degree SI). (b) Normalized infection  as a function of time. Blue line for using the test-kits in node C and D . Green line for applying test-kit in A and B. Infection is significantly reduced for the choices of green patches. (c) Normalized  final outbreak size ($\mathcal{Z}_{\rm FOS}$) as a function of $\beta$ (disease transmission rate). The outbreak size decreases for SW. The black lines in (b) and (c) reveal the evaluation of $\mathcal I$ and $\mathcal{Z}_{\rm FOS}$ in absence of any test-kits. }  
	\label{Fig:small_network_I_Z_FOS}
\end{figure}
To illustrate our strategy, we choose  a  small but heterogeneous network (Fig.\ \ref{Fig:small_network_I_Z_FOS}(a)) of 
$40$ nodes (patches).  To identify the  influential patches in the network, we have chosen four nodes with different degrees marked by $A,B,C$ and $D$.    The patches 
$A$ ($d_A=19$) and $B$ ($d_B=9$) (green) have large number  of neighbours  where as  
$C$ ($d_C=5$) and $D$ ($d_D=4$) 
(blue) have less number of connections. Rest of the  nodes (grey) have  small number of connections.  
At first we apply the test-kits in $A$ and $B$  
following the {\it{degree based strategy}} (SD) e.g. the test-kits will be  distributed and  divided according to the degree of those patches.  Therefore, the test-kit will appear in two equations  only with the term $p_A=\frac{d_A}{d_A+d_B}\sim 0.678 $ 
and $p_B=\frac{d_B}{d_A+d_B} \sim 0.321$, respectively.
Next, we have randomly chosen two low degree patches $C$ and $D$ (blue circles in Fig.\ \ref{Fig:small_network_I_Z_FOS}(a)) and 
distribute the test-kit equally, i.e., 
$p_C=\frac{1}{2}$ and $p_D=\frac{1}{2}$. 
We call it as {\it{ randomly selected and  identically distributed  test-kit strategy}} (SI). These two  strategies have different impact in the evaluation of total infection 
($\mathcal I$). 
Compared to SI (blue line in Fig.\ \ref{Fig:small_network_I_Z_FOS}(b)),  the peak of the infection is significantly decreased for SD (green line in  Fig.\ \ref{Fig:small_network_I_Z_FOS}(b)). In absence of  test-kits, the total infection
($\mathcal I_{\rm SW}$) 
is shown with black line (strategy without kit: SW) in the same figure.  To quantify the efficiency of each strategy, we define   relative reduction 
($RR$) of the peak of infection as follows:
\begin{eqnarray}
{ RR_{SI/SD}}= \frac{\mathcal{I_{\rm SW}}-\mathcal{I_{\rm SI/SD}}}{\mathcal{I_{\rm SW}}}\times 100\%.
\end{eqnarray}
From the Fig.\ \ref{Fig:small_network_I_Z_FOS}(b), it is clear  $RR_{SI} \approx25\%$ 
and $RR_{SD} \approx 70\%$ 
ensuring the efficiency of degree based strategy (SD). We  have also plotted the   final outbreak size 
($\mathcal{Z}_{\rm FOS}$) 
as a  function of disease transmission rate $\beta$ shown in Fig.\ \ref{Fig:small_network_I_Z_FOS}(c).  The
$\mathcal{Z}_{\rm FOS}$  
is significantly decreased for degree based strategy (SD, green) with respect to SI (blue)  and SW (black). Here the migration strength is fixed at 
$\epsilon=0.05$.
Note that, there will be several ways to choose specific patches from the network. For instance, we can choose patches 
$A$ and $C$ or $B$ and $D$. 
However these choices will not be effective compared to the choices of $A$ and $B$. 
We have now guessed that for a  suitable migration  strength, applying test-kits in the high degree patches  will be  highly beneficial for connected communities. A natural question appears, what will happen if we choose other microscopic topological properties of a given graph?   { Apart from the degree sequence, here we introduce two more  network characteristics: (i)  Betweenness centrality \cite{newman2003structure,newman2018networks,boccaletti2006complex}, and (ii) local clustering coefficient \cite{newman2003structure,watts1998collective,boccaletti2006complex}. For definition and detailed description of these network characteristics, please see the supplementary material, Sec.\ \ref{Comparison of   diverse centrality measures with degree}.  We will call them as SB and SC based strategy.
 Now we check our test-kit approach for three local structural  measures:  degree, betweenness and clustering}.

\begin{figure}
	\includegraphics[width=1\textwidth]{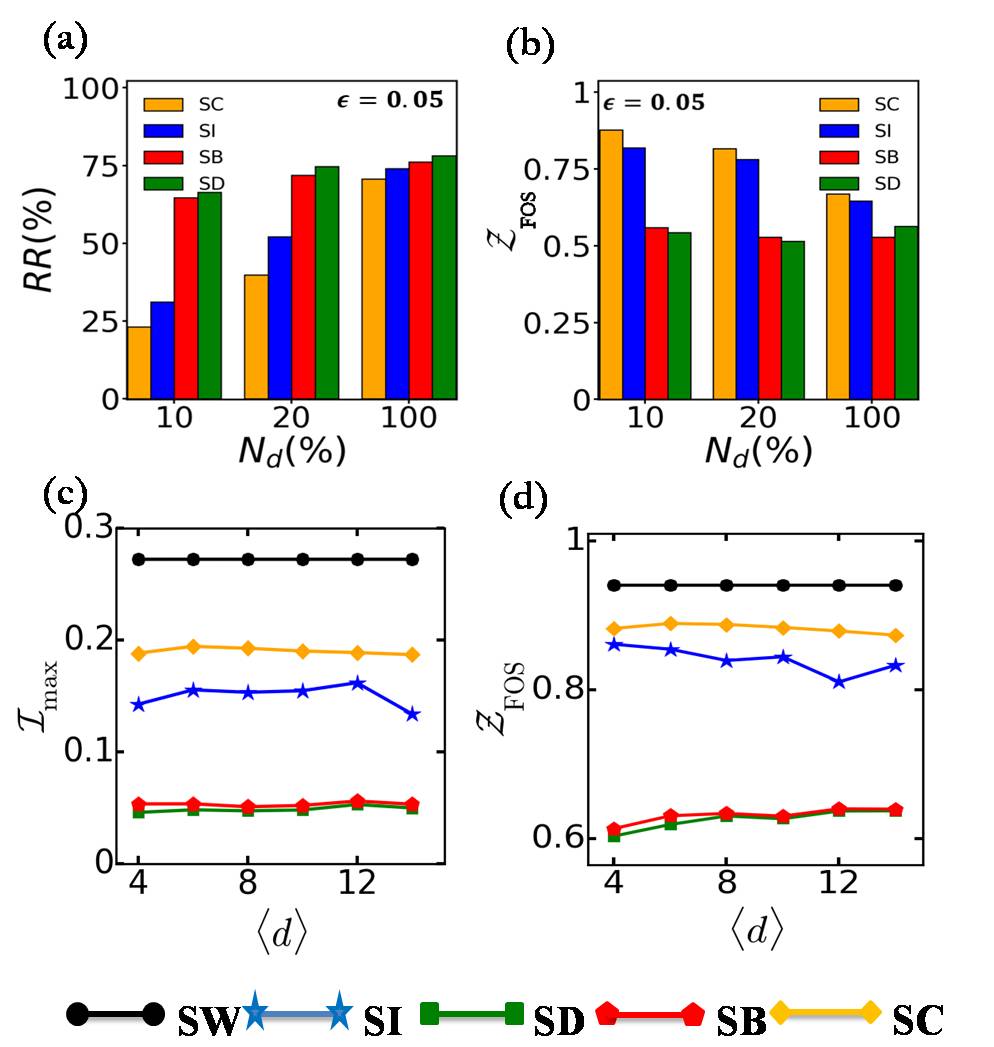}
	\caption{ {\bf Impact of  network matrices and densities on relative reduction ($RR$) of infection peak and final outbreak size ($Z_{\rm FOS}$)}. (a) Four strategies are chosen: degree based (SD, deep green bar), clustering coefficient base (SC, orange), betweenness  centrality based (SB, red) and randomly chosen but identically distributed (SI,blue). For SD, SC, and SB,  $10\%$ as well as $20\%$ patches with higher degrees, clustering coefficients and betweenness centralities are chosen. Noticeably, SD and SC have more capability of reducing the peak of infection. The results are equally  improved for all cases if we apply test-kits in all of the patches ($100\%$ extreme right part). (b) SD and SB can significantly decrease the outbreak size for all cases (compared to SI, SC). (c-d) Impact of average degree $\langle d \rangle$ on $\mathcal{I}_{\rm{max}}$ and $Z_{\rm FOS}$. Network size is fixed $M=500$, and each network is scale free by nature. Here $\beta=0.03$. All the other parameters are taken same as in Fig.\ \ref{single_model_plot}. }  
	\label{Fig:Barplot}
\end{figure}

In our work, we use a  heterogeneous scale-free network of size $M=500$ with average degree 
$\langle d \rangle=14$ 
and exponent $3$. 
At first  we apply test-kits in a fraction of nodes 
($N_d=10\%$) 
which  have (i) larger degrees, (ii) larger clustering coefficients, or (iii) larger node betweenness  centralities compared to the rest of the nodes. The results are shown in  the extreme left  of Fig.\ \ref{Fig:Barplot}(a) for relative reduction ($RR$) of infection peak and (b) for final outbreak size.  It is clear that, the infection peak  is significantly reduced ($\sim 65\%-70\%$)  
for degree based (SD, green bar) as well as for betweenness centrality based strategy (SB, red bar). On the other hand, SI and SC (blue and orange bar respectively) based strategy can only reduce $25\%-30\%$ of the peak of infection. We have observed the similar pattern for the final outbreak size (extreme left Fig.\ \ref{Fig:Barplot}(b)) where 
$\mathcal{Z}_{\rm FOS}$ 
is remarkably  decreased for SB and SD strategy. If we increase the  fraction of nodes to $N_d=20\%$, the situation is slightly improved for  SI and SC based strategy, however they cannot outperform SB and SD.
When we apply test-kits in every nodes
($100\%$, extreme right in each panel), 
all the strategies provide almost similar results.
 To check whether our partial control based strategies are robust, we increase the average degree 
($\langle d \rangle$)
of the network ($M=500$) 
continuously from 
$4$ to $14$. The impact of increased average degree is 
shown in Fig.\ \ref{Fig:Barplot}(c-d). The peak of the total  infection 
$\mathcal{I}_{\rm{max}}$ 
is reported  for four different strategies: degree based (SD, green),  betweenness centrality based (SB, red),  clustering and randomly chosen identically distributed  (SC, SI in orange, blue respectively).  For comparison, we have also shown the infection peak in absence of test-kit (black line (SW)).   Clearly, the results are not affected by the increased average degree of  the network, i.e., SD and SB become the suitable choice for test-kit distributions. Same thing occurs for final outbreak size, the prevalence is reduced and remain almost constant (around $0.6$) for SD and SB.
For  Fig.\ \ref{Fig:Barplot}(c-d), the $10\%$  nodes/patches are impacted by the test-kits, and coupling strength $\epsilon$ is fixed at $0.05$. { Changes in $RR$ with respect to $N_d$ and $\epsilon$ have been explored in the next section}.   

\subsection{Simultaneous impact of $\epsilon$ and $N_{d}$ on $\mathcal{I}_{\rm max}$ and $\mathcal{Z}_{\rm FOS}$ for different  strategies}
\begin{figure}
	{\includegraphics[width= 1\textwidth]{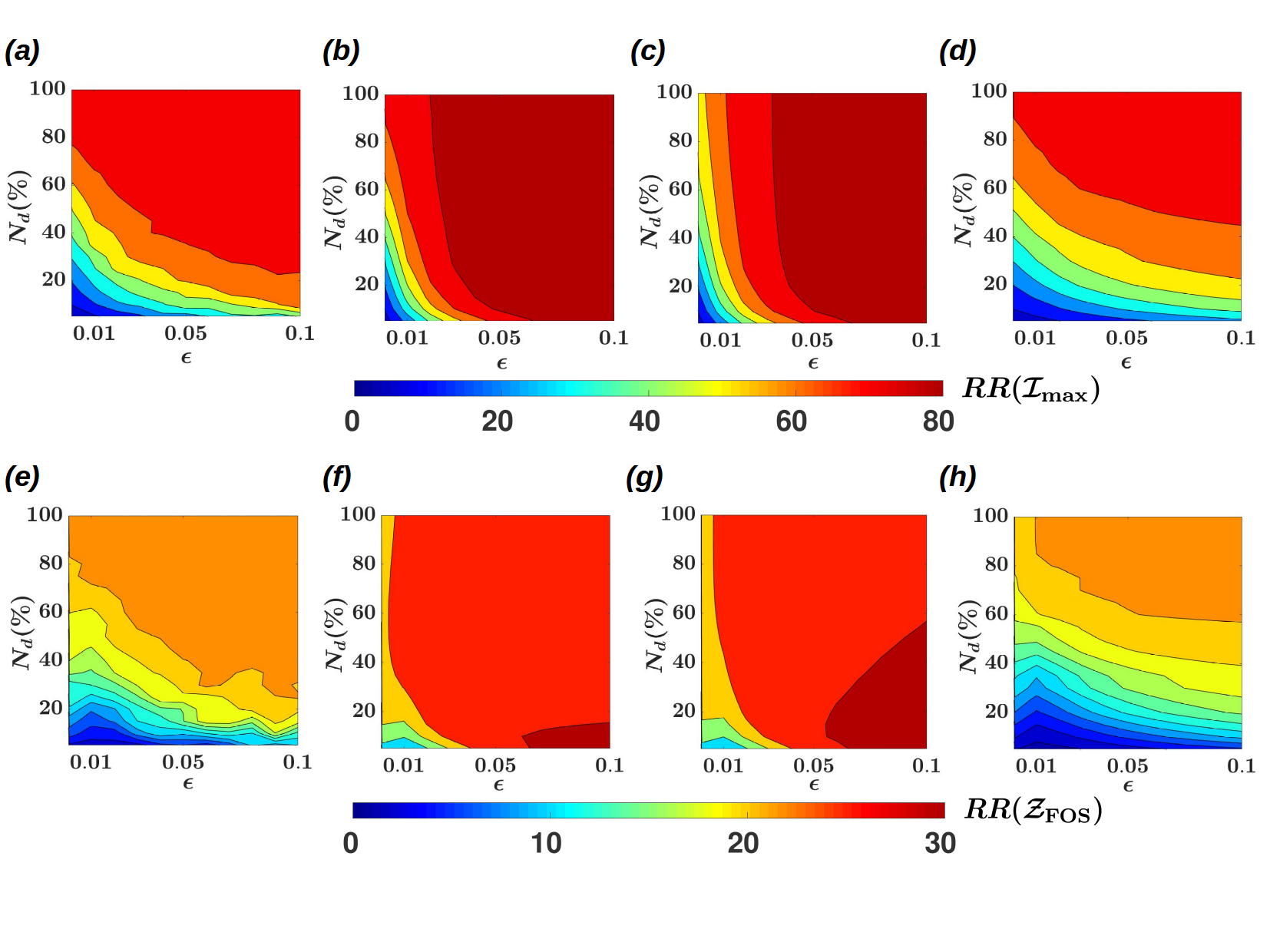}}
	\caption{(a)-(d) Percentage of relative reductions ($RR$) in 
		$\mathcal{I}_{\rm max}$ 
		for the four different strategies SI, SD, SB, and SC respectively by varying both coupling strength ($\epsilon$) 
		and percentage of controlled patches ($N_{ d}$) 
		simultaneously. In relatively higher coupling strength, $\epsilon \sim~0.05$, SD and SB can reduce $\mathcal{I}_{\rm max}$ approximately $80\%$ even if very low number of nodes are targeted. (e)-(h) Percentage of relative reductions ($RR$) in $\mathcal{Z}_{\rm FOS}$ for the four different strategies SI, SD, SB, and SC respectively by varying both coupling strength ($\epsilon$) and percentage of controlled patches ($N_{ d}$) simultaneously.  Here $\beta=0.03$. The rest of the  parameters are taken same as in Fig.\ \ref{single_model_plot}.    }      
	\label{fig:sm_1}
\end{figure}

Here, we investigate the efficiencies of four strategies: SI, SD, SB, and SC for different coupling strength
($\epsilon$)
and for different percentage of targeted nodes 
($N_{d}$). 
To assess the efficiency of a particular strategy X (=SI, SD, SB, SC), we define percentage of relative reduction in $\mathcal{I}_{\rm max}$ ($RR^{\rm X}(\mathcal{I}_{\rm max})$) 
and percentage of relative reduction in 
$\mathcal{Z}_{\rm FOS}$ ($RR^{\rm X}(\mathcal{Z}_{\rm FOS})$) as follows:
\begin{eqnarray}
RR^{X}(\mathcal{I}_{\rm max})=\frac{\mathcal{I}_{\rm max}^{\rm SW}-\mathcal{I}_{\rm max}^{\rm X}}{\mathcal{I}_{\rm max}^{\rm SW}}\times 100\%, {~\rm ~~~ ~}\nonumber
RR^{X}(\mathcal{Z}_{\rm FOS})=\frac{\mathcal{Z}_{\rm FOS}^{\rm SW}-\mathcal{Z}_{\rm FOS}^{\rm X}}{\mathcal{Z}_{\rm FOS}^{\rm SW}}\times 100\%,
\end{eqnarray}
where, $\mathcal{I_{\rm max}^{\rm SW}}$ 
and $\mathcal{Z}_{\rm FOS}^{\rm SW}$ 
are the peak value of infection   and final outbreak size in absence of test-kit (SW), respectively. From Figs.\ \ref{fig:sm_1}(a)-(d), we see that for each strategy, $RR(\mathcal{I}_{\rm max})$
gradually increases with the increment of 
$\epsilon$ 
and 
$N_{d}$.
Now if we compare among these four strategies, we see that for the strategies: SD and SB, for coupling strength, 
$\epsilon \sim 0.05$, $RR(\mathcal{I}_{\rm max})$ 
reaches approximately $80\%$ for a very low number of targeted patches (see Fig.\ \ref{fig:sm_1}(b)-(c), deep red). 
On the other hand, for the strategies SI and SC,  for the higher coupling strength and lower percentage of targeted nodes (even higher fraction of targeted nodes) such high percentage 
($\sim 80\%$) of reduction in
$\mathcal{I}_{\rm max}$
is not possible (see Fig.\ \ref{fig:sm_1}(a),(d), red color). 
\par  
Next we investigate the efficiency in terms of $RR(\mathcal{Z}_{\rm FOS})$ 
for the four above-mentioned strategies. In this case, we also observe a similar trend as in 
$RR(\mathcal{I}_{\rm max})$, 
i.e., relative reductions in 
$\mathcal{Z}_{\rm FOS}$ increase with the increment of 
$\epsilon$ as well as of $N_{d}$ 
(see Fig.\ \ref{fig:sm_1} (e)-(h)). 
In the same way,  it is to be noted that in terms of $RR(\mathcal{Z}_{\rm FOS})$,
the strategies SD and SB appear to be quite efficient than SI and SC. 
The strategies SI and SC can achieve maximum $20\%$ reduction (see Fig.\ \ref{fig:sm_1}(e) and (h), yellow and light yellow regime) in $\mathcal{Z}_{\rm FOS}$, whereas  $30\%$ reduction can be achieved through the strategies SD and SB (see Fig.\ \ref{fig:sm_1}(f)-(g), red or deep red regime).       
 In our cases, it is clear, degree based and betweenness based  strategy (SD and SB) outperform the others.  The reason is as follows: 
 In heterogeneous network, removal of most connected  nodes can break the giant component into isolated fragments \cite{albert2000error}. Also, it was shown that isolating small subset of nodes (with high connectivity) for vaccination may resist the epidemic outbreak \cite{eubank2004modelling,madar2004immunization, chen2008finding,liu2016biologically,tanaka2014random,bucur2020beyond}.   As high degree nodes have large number of connections, it helps to spread  information more rapidly in the entire graph.  Therefore, targeting few nodes (high degrees) at the onset of epidemics and  apply test-kits in the targetted patches the prevalence become less severe, therefore a large number of neighbours got less infected. The same thing occurs for the case of  betweenness centrality.
However, the clustering based  strategy (SC) which signifies how the neighbours of  a node  are connected to each other, cannot encode the information of the coupling structure, therefore fails to improve the result. Same thing occurs for randomly chosen but identically distributed strategy (SI), since the chosen nodes and their neighbor cannot capture the underline heterogeneous structure at all. In  the same way, we can identify   the nodes with high page rank or closeness characteristics which will may have positive impact in the intervention strategy. We have elaborately this issue in the  supplementary material (SM) Sec. \ref{Comparison of   diverse centrality measures with degree}, Fig.\ 9. 

\subsection{Impact of transmission rate $\beta$ on $\mathcal{I}_{\rm{max}}$ and $\mathcal{Z}_{\rm{FOS}}$}
\begin{figure*}
	\includegraphics[width=1\textwidth]{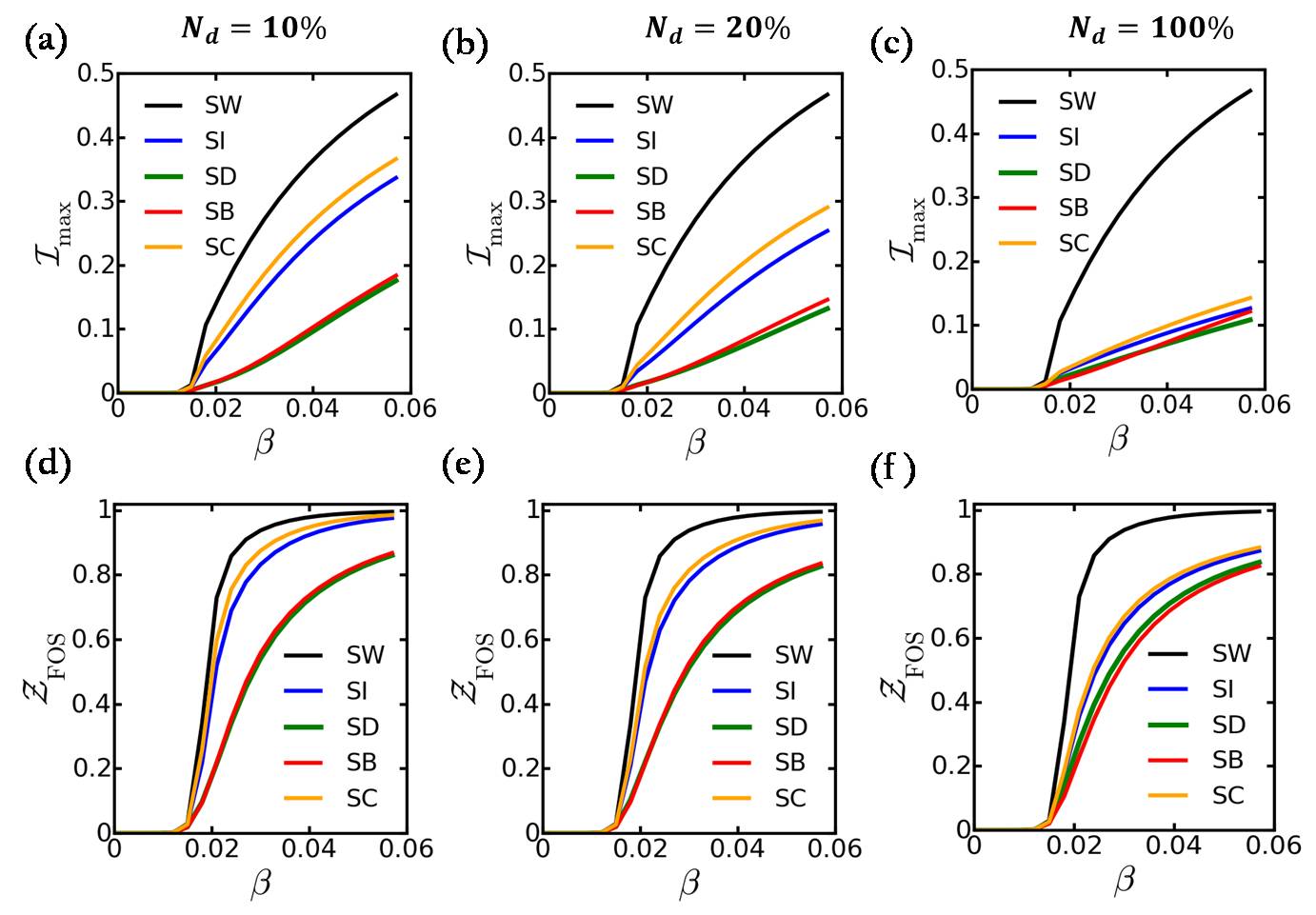}
	\caption{{\bf Normalized Cumulative infection peak and final outbreak size vs $\beta$.} (a)(d) Testing kits are provided in $10\%$ patches. At higher transmission rate  ($\beta\sim 0.06$) the $\mathcal{I_{\rm{max}}}$ 
		is approximately $0.45$ for SW (black line), where as $\mathcal{I_{\rm{max}}}\sim 0.15$  for SD and SB (red, green). The clustering based strategy (SC, orange) and strategy based on identically distributed test-kits in randomly chosen patches (SI, blue) are relatively small than  the  SW and higher than SB, SD. Here $\mathcal{Z}_{\rm{FOS}}$ 
		is saturated 
		($\beta\sim 0.06$) 
		around $1$ for SW, SI and SC where as $\mathcal{Z}_{\rm{FOS}}\sim 0.8$ 
		for SD, SB. (b) (e) Now $N_d$ is increased to $20\%$. SC and SI are slightly improved for both cases. (c) (f) $N_d=100\%$, all the strategies have similar impact on 
		$\mathcal{I_{\rm{max}}}$. 
		However for $\mathcal{Z}_{\rm{FOS}}$ 
		the SD and SB are still better than  SI and SC. The coupling is fixed at $\epsilon=0.05$.  In the previous figures $\beta$  is fixed at $0.03$. All the other parameters are taken same as in Fig.\ \ref{single_model_plot}.  }  
	\label{Fig:beta_vsI_max_ZFOS_plot}	
\end{figure*}

{We have established now, in an heterogeneous network,  degree-based or betweenness based test-kit strategy has strong ability to reduce the infection peak as well as final outbreak size. 
	To delve deeper, we have further cross validated our proposed for a wide range of infection rate ($\beta$). Now we will show,  in absence of test-kit  increased $\beta$  enhances the peak infection significantly. On the other hand, partial implication of  test-kits slows down the infection peak ($\mathcal{I}_{\rm{max}}$) 
as well as  final outbreak size ($\mathcal{Z_{\rm{FOS}}}$)  for a wide range of $\beta$. }
Particularly,  when a  small fraction of patches (with high betweenness or degree) are monitored  by the test-kits ($N_d=10\%$), 
the 
$\mathcal{I}_{\rm{max}}$ 
is significantly decreased shown in    Fig.\ \ref{Fig:beta_vsI_max_ZFOS_plot}(a) for SD (green) and SB (red). 
In higher  transmission rate ($\beta\sim 0.06$)
the peak of infection are  enhanced for SW (black), SI (blue) and for SC (magenta). However, SB and SD based strategy largely restrict ($\sim 0.15$)  the peak of infection (red and green) around.  A slight increase in the number of controlled patches 
($N_{d}=20\%$) 
improves the decrement of $\mathcal{I}_{\rm{max}}$ for strategy SC and SI shown in Fig.\ \ref{Fig:beta_vsI_max_ZFOS_plot}(b). In this situation, the  impact of SD and SC are still superior than SI and SC. If we apply test-kits in all patches($N_d=100\%$), the output of  all strategies are almost same irrespective of the value of the transmission parameter $\beta$ (Fig.\ \ref{Fig:beta_vsI_max_ZFOS_plot}(c)).   Similar type of patterns appear for the case of $\mathcal{Z}_{\rm{FOS}}$ shown in  Fig.\ \ref{Fig:beta_vsI_max_ZFOS_plot}(d-f) where 
$N_d=10,20, {\rm and} ~ 100\%$ 
respectively.  For all $N_d$, the SD and SB based strategy outperform the other two strategies.  Note that, for all cases described above (Fig.\ \ref{Fig:beta_vsI_max_ZFOS_plot}(a-f)), the  onset of infection 
($\mathcal{I}>0$) starts around 
$\beta \approx 0.012$. 
{
In the appendix we have analytically calculated  the critical transmission rate  ($\beta_c$) in the network and identified that 
$\beta_c=0.01$  for $\mathcal{R}_{0}^{\rm network}>1$ and $\alpha_0=0.01$.  Interestingly, the critical $\beta$ is same as single model and it does not depend on the network structure as well as diffusive coupling strength $\epsilon$}. 

\section{Optimal test-kit based strategy in real networks}
For further   validation of optimal  test-kit based strategy,  we have considered two real networks: one is a connectivity pattern of international airports  through flights and the other is a transportation network  within  Wards of Kolkata municipality corporation. 
\par
\begin{figure*}[h]
	{\includegraphics[width= 1\textwidth]{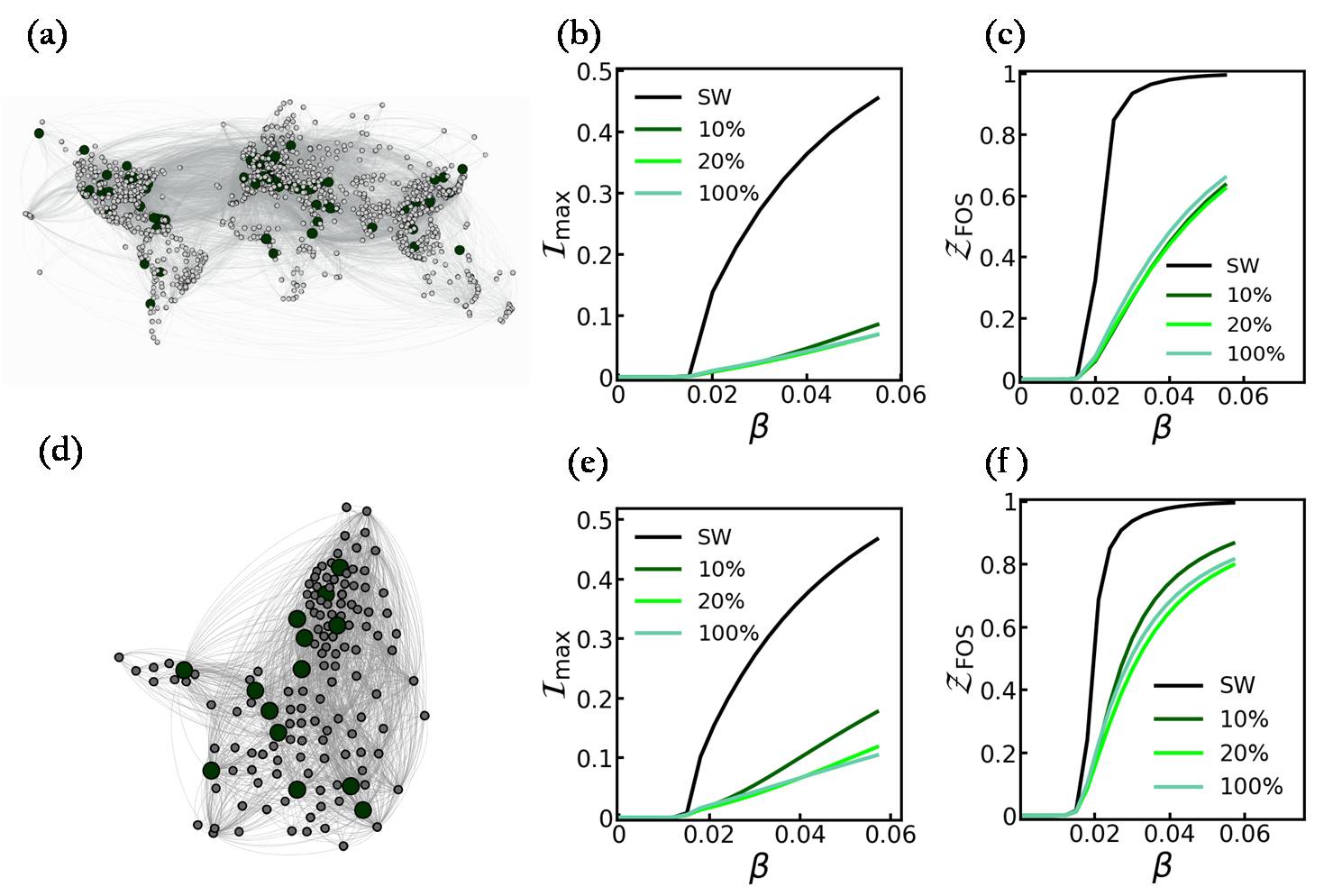}}
	\caption{\textbf{$\mathcal{I}_{\rm max}$ and $\mathcal{Z}_{\rm FOS}$ vs $\beta$ on real networks.} (a) Airport network. $10\%$ high degree nodes are marked with deep green circle. (d) Kolkata transportation network. (b,e) $\mathcal{I}_{\rm max}$ as a function $\beta$. The black line (SW) represents the infection {peak} in absence of test kits. Deep green line is drawn when test-kits are applied at $10\%$ high degree nodes.  Light green lines are drawn when  test-kits are  applied to the $20\%$  high degree nodes. The result remain almost same for  $100\%$ intervention (lighter green).  (c,f)  Final outbreak size as a function $\beta$. Partial control of network has the same impact as compared to the full control of the network. All the other parameters are taken same as in Fig.\ \ref{single_model_plot}. }
	\label{fig:sm_3}
\end{figure*}

 A global airport network of nodes $1292$  linked
through $38, 377$ directional air-routes \cite{hens2019spatiotemporal,brockmann2013hidden} is considered.  The network is shown in the Fig.\ \ref{fig:sm_3}(a). A fraction  ($10 \%$) of nodes wth higher degree are identified (deep green circles in the network) and test-kits are provided only in these airports. Compared to the absence of testing kits, the $\mathcal{I}_{\rm max}$ and $\mathcal{Z}_{\rm FOS}$ are significantly decreased if we apply test-kit in  $10 \%$ high degree nodes (deep green line Fig.\ \ref{fig:sm_3}(b-c)). Qualitatively, the results are  not changed if we increase the number of controlled nodes (light green lines). The migration strength is fixed at $0.05$. 
However, our result is valid for a wide range of migration strength (the results are not shown in this work).
\par Next,  we construct Kolkata transportation network. The network covers the links due to the connectivity patterns of major bus routes, the metro connection within the city. We have also added links between adjacent Wards to mimic the mobility of  pedestrians as well as the movements of three-wheeler  passenger cars (auto rickshaw).   The network has $141$ nodes (Wards) and $2524$ edges. { For more details please see the Supplementary material, Sec.\ \ref{Description of real networks}.}
Here we have also identified $10\%$ nodes having high degrees (Fig.\ \ref{fig:sm_3}(d), deep green).   A partial  control (intervention of testing-kits) in a certain fraction of Wards significantly decrease the peak infection as well as the final outbreak size (Fig.\ \ref{fig:sm_3}(e-f)).{ The optimal control strategy has same impact as in airport network as well as the synthetic network described earlier.}.

\section{Conclusion}
Our  compartmental model reveals the suitable implementation of testing kits can reduce severe effect of epidemics.  {A rigorous analytical derivation associated with numerical simulation confirms our claims. We have confirmed that such strategy can reduce (i) the peak of the infection as well as the (ii) the final outbreak size. To validate our model, we have   randomly chosen three states of USA and confirmed that our model can efficiently capture the time dependent hospital data of each state.}  Next,  we have  considered the  interlinked  meta-communities and tested the impact of  the migration (between communities/patches) and nodal characteristics on the underlying networks. {To perform the numerical simulation, the proposed   model is used over the top of a  heterogeneous graph and connect each patch with diffusive coupling.  The migration of people from one patch to other patch is considered through diffusive coupling reflecting intrinsic functional activities of meta-communities.}  In this backdrop, we have partially intervened the networks with test-kits, e.g the test-kits are applied in the nodes which  have higher degrees or betweenness centralities. The results reveal that suitable choices of the nodes for the test-kit implication can have same effect compared to the intervention of test-kits in the entire graph. To validate our  networks results further,    we have considered two real networks (i)  airport to airport connectivity pattern through existing flights and (ii) transportation network within wards in  Kolkata municipality.  We have numerically confirmed that, in an environment of  a degree based strategy, compared to full control, a partial intervention ($10\%$ nodes are chosen for the test-kit implications) can drastically  reduce the final outbreak size as well as infection peak.

\section{ Appendix: Derivation of basic reproduction number ($\mathcal{R}_{0}^{\rm network}$) in heterogeneous network }
Here we derive the basic reproduction number for our network model (see the Eqn.\ (4.1)). Following the same procedure as described for single model, we calculate the new infection matrix $\mathcal{F}$ and transmission matrix $\mathcal{V}$ as follows:
\[
\mathcal{F}=
\left[{\begin{array}{cc}
	\mathbb{O}_{M\times M} & \mathbb{F}_{12}\\
	\mathbb{O}_{M\times M} & \mathbb{O}_{M\times M}\\
	\end{array}} \right]_{2M\times 2M}
\] 
where 
\begin{eqnarray} 
\nonumber 
\mathbb{F}_{12} &=& \rm diag(\beta_{n}), n=1,2,...,M,
\label{F_12}      
\end{eqnarray}
and 
\[
\mathcal{V}=
\left[{\begin{array}{cc}
	\mathbb{V}_{11}& \mathbb{O}_{M\times M}\\
	-\mathbb{V}_{21} & \mathbb{V}_{22}\\
	\end{array}} \right]_{2M\times 2M}
\]
where,
$\mathbb{V}_{11} = \rm diag(\sigma + \epsilon) - \tilde A$, $\mathbb{V}_{21} = \rm diag(\sigma)$, $\mathbb{V}_{22} = \rm diag(\alpha_{0} + \epsilon) - \tilde A$ and the elements of $\tilde A$ is given by $\tilde A_{ij}=\epsilon \frac{A_{ij}}{d_{i}}$, $A_{ij}$ denotes the elements of adjacency matrix and $d_{i}$ is the degree of $i^{th}$ node.
\par Since $\mathbb{V}_{11}$ and $\mathbb{V}_{22}$ both are irreducible non-singular M-matrices with positive column sums, we have $\mathbb{V}_{11}^{-1}>0$ and $\mathbb{V}_{22}^{-1}> 0$~\cite{hsieh2007impact}.

Now $\mathcal{V}^{-1}$ can be written as 
\[
\mathcal{V}^{-1}=
\left[{\begin{array}{cc}
	\mathbb{V}_{11}^{-1} & \mathbb{O}\\
	\mathbb{V}_{22}^{-1}\mathbb{V}_{21}\mathbb{V}_{11}^{-1} & \mathbb{V}_{22}^{-1}\\
	\end{array}} \right].
\] 
Finally, $\mathcal{F}\mathcal{V}^{-1}$ can be expressed as 
\[
\mathcal{F}\mathcal{V}^{-1}=
\left[{\begin{array}{cc}
	\mathbb{F}_{12} \mathbb{V}_{22}^{-1}\mathbb{V}_{21}\mathbb{V}_{11}^{-1} & \mathbb{F}_{12} \mathbb{V}_{22}^{-1}\\
	\mathbb{O} & \mathbb{O}\\
	\end{array}} \right]
\] 
The basic reproduction number for the network model~\eqref{eq:network_model}, $\mathcal{R}_{0}^{\rm network}$ is given by,
\begin{eqnarray}  
\mathcal{R}_{0}^{\rm network}&=& \rho (\mathbb{F}_{12} \mathbb{V}_{22}^{-1}\mathbb{V}_{21}\mathbb{V}_{11}^{-1} ),
\label{R_0_n}      
\end{eqnarray}
where $\rho(A)$ 
denotes the spectral radius of the matrix $A$.
\par Since $\mathbb{V}_{21}$ 
is a diagonal matrix and in our study all the transmission rates are taken same, i.e., $\beta_{n}=\beta$, 
for $n=1,2,...,M$, 
the expression of 
$\mathcal{R}_{0}^ {\rm network}$ 
can be further simplified as:
\begin{eqnarray} 
\mathcal{R}_{0}^{\rm network}=\beta \sigma \rho\big((\mathbb{V}_{11}\mathbb{V}_{22})^{-1}\big).
\label{eq:R0_n_1}
\end{eqnarray}
Now we find analytical expression of  $\rho(\mathbb{V}_{22}^{-1}\mathbb{V}_{11}^{-1})$ by using the properties of the matrices $\mathbb{V}_{11}$ and $\mathbb{V}_{22}$. We first note that, the matrices $\mathbb{V}_{11}$ and $\mathbb{V}_{22}$ are basically a translation of the matrix $-\tilde A$ by a scalar multiple of identity matrix. Therefore, it is very easy to check that if $v$ is an eigenvector corresponding to the eigenvalue $\lambda$ of the matrix $-\tilde A$, the matrices $\mathbb{V}_{11}$ and $\mathbb{V}_{22}$  will have the same eigenvector corresponding to the eigenvalue $\sigma + \epsilon + \lambda $ and $\alpha_{0} + \epsilon + \lambda$ respectively. 
\par 
Since, $\tilde A$ is a non-negative matrix, and let $\lambda_{\rm max}$ be the maximum eigenvalue of the matrix $\tilde A$, then the bound on $\lambda_{\rm max}$ can be obtained using Perron-Frobenius inequality \cite{maccluer2000many} as follows,
\begin{center}
	${\rm min}_i(\epsilon \sum_{j=1}^{\mathrm{M}} \frac{A_{ij}}{d_{i}}) \leqslant \lambda_{\rm max} \leqslant {\rm  max}_{i}( \epsilon \sum_{j=1}^{\mathrm{M}} \frac{A_{ij}}{d_{i}}).$    
\end{center}
Again,  $\sum_{j=1}^{\mathrm{M}} \frac{A_{ij}}{d_{i}}=1$, 
we end up with 
$\lambda_{\rm max}=\epsilon$. 
Therefore, the minimum eigenvalue of the matrices 
$\mathbb{V}_{11}$ ($\mathbb{V}_{11} = \rm diag(\sigma + \epsilon) - \tilde A$) and $\mathbb{V}_{22}$ ($\mathbb{V}_{22} = \rm diag(\alpha_{0} + \epsilon) - \tilde A$) 
will be  $\sigma$ and $\alpha_{0}$ respectively. As we have shown, $\mathbb{V}_{11}$ and $\mathbb{V}_{22}$  share common eigenvector and commute to each other, it is straight forward to check that the maximum eigenvalue of the matrix $(\mathbb{V}_{11}\mathbb{V}_{22})^{-1}$ 
is $\frac{1}{\alpha_{0}\sigma}$. 
Using these mathematical arguments in Eqn.\ \eqref{eq:R0_n_1}, we finally get the mathematical expression of $\mathcal{R}_{0}^{\rm network}$as,
\begin{eqnarray}
\mathcal{R}_{0}^{\rm network}=\frac{\beta}{\alpha_{0}}.
\label{eq:R0_network_final}
\end{eqnarray}

\newpage
\begin{center}\Large{Supplemental Material for: ``Optimal test-kit based intervention strategy  of epidemic spreading in heterogeneous complex networks''}\end{center}
\author{Subrata Ghosh$^{1}$}
\thanks{Equal contribution}
\author{Abhishek Senapati$^{2}$}
\thanks{Equal contribution}
\author{Joydev Chattopadhyay$^{2}$}
\author{Chittaranjan Hens$^{1}$}
\thanks{Corresponding Author}
\email{chittaranjanhens@gmail.com}

\author{Dibakar Ghosh$^{1}$}

\affiliation{\noindent \textit{$^{1}$Physics and Applied Mathematics Unit, Indian Statistical Institute, 203 B. T. Road, Kolkata 700108, India}}
\affiliation{\noindent \textit{$^{2}$Agricultural and Ecological Research Unit,, Indian Statistical Institute, 203 B. T. Road, Kolkata 700108, India}}
\maketitle
\section{Best-fit parameters}
\label{Best-fit parameters}
 We estimate four unknown model parameters: (i) the disease transmission rate ($\beta$), (ii) effectiveness of test-kit ($\alpha_{1}$), (iii) rate of production of kit ($\xi$), (iv) rate of losing efficacy of test-kit ($\chi$) by fitting our model to cumulative daily COVID-19 hospitalized data for three states of U.S.A. At any time instant $t$, the cumulative number of hospitalized persons from the model is given by Eq. 3.1 in the main Text. The model fitting is executed by minimizing the sum of square function $SS(\Theta)$ (see Eq. 3.2 in Main Text) using in-built function \textit{lsqnonlin} in MATLAB (Mathworks, R2014a). The values of the \textit{best-fit} parameters ($\hat{\beta},\hat{\alpha_{1}}, \hat{\xi}, \hat{\chi}$) are given in Table~\ref{table:fitted_parameters}.

\begin{table}[h!]
	\tabcolsep 7pt
	\centering
	\begin{tabular}{p{3cm} p{1cm} p{1cm} p{1cm} p{1cm}}
		\hline\hline
		\textbf{State} & \boldmath$\hat{\beta}$ & \boldmath$\hat{\alpha_{1}}$ &\boldmath$\hat{\xi}$& \boldmath$\hat{\chi}$ \\
		[0.5ex]
		\hline
		
		\small{Maryland} & \small{0.6591} & \small{0.0298} & \small{0.0032}&\small{0.0180} \\\\
	    \small{Ohio} & \small{0.6195} & \small{0.1035} & \small{0.0025}&\small{0.0630} \\\\
		\small{New York} & \small{0.7690} & \small{0.0015}&\small{0.0076}& \small{0.0001} \\\\
		\hline
	\end{tabular}
	\caption{Best-fit parameter values obtained from non-linear least square method for the three states of U.S.A.}
	\label{table:fitted_parameters}
\end{table}

\section{Random vs. targeted control}
\label{Random vs. targeted control}
\begin{figure}
	{\includegraphics[width= 1\textwidth]{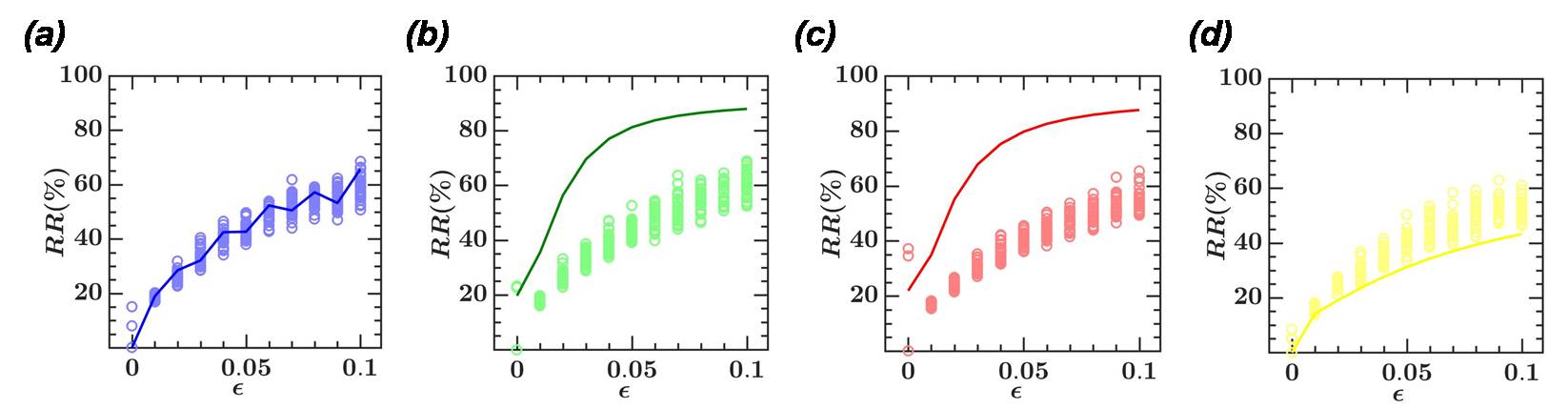}}
	\caption{\textbf{Random control vs. targeted control.} (a)-(d) Relative reduction in $\mathcal{I}_{\rm max}$ ($RR$) with respect to coupling strength $\epsilon$ for the strategies SI, SD, SB, SC respectively. For each case, solid line represents the results corresponding to targeted control and the small circles denotes the results corresponding to different realizations of random control for a given coupling strength $\epsilon$. The number of realization for random control is $30$.  }
	\label{fig:sm_2}
\end{figure}

We now compare our proposed controlled strategy by  selecting another set of nodes. In that case, the nodes  are randomly  chosen instead of targeting the nodes bearing the respective higher network measures (i.e degree, betweenness and clustering coefficients). Note that, for each strategy, after selecting the random nodes we employ the test-kits according to the proportion of their network scores. However for SI, the test-kits are always distributed identically. 
To study the impact of randomization in each strategy, we fix the $N_{d}$ (controlled nodes/patches) at 
$10\%$ 
and quantify the relative reduction in 
$\mathcal{I}_{\rm max}$ i.e $RR$ 
for a wide range of coupling strength, $\epsilon$. For all the strategies, we plot $RR$ 
obtained for targeted control (nodes having high network scores and test-kits are distributed according to their scores: degree, betweenness or clustering) and random control (randomly selected nodes and test-kits are distributed according to their network score) by varying the coupling strength $\epsilon$. 
For random control we take $30$ different realizations. Since in the strategy SI, all the nodes have same status (i.e $p_{n}=\frac{1}{M}$, for $n=1,2,...,M$), we observe no such significant difference between the targeted control and random control (see Fig.\ \ref{fig:sm_2}(a), blue line and blue circle respectively). However, in the cases of SD and SB, a significant difference between targeted control and random control is observed. From Fig.\ \ref{fig:sm_2}(b)-(c), we see that in terms of reduction in $\mathcal{I}_{\rm max}$, random control (green and red circles) performs poorly in comparison to targeted control (green and red lines) for any given coupling strength. For the strategy SC, we see that random control (see Fig.\ \ref{fig:sm_2}(d), yellow circles) gives slightly better results than targeted control (yellow line) in reducing $\mathcal{I}_{\rm max}$. 

\section{Comparison of   diverse centrality measures with degree}
\label{Comparison of   diverse centrality measures with degree}
We have already shown that fraction  of nodes having large connections or  larger between centralities can significantly reduce the outbreak size. However clustering coefficient based strategy cannot do that. 
 Apart from the degree sequence,  here we revisit  the  other structural  properties of nodes  to understand their correlation with degree sequence. In our case, we select:  (i)  Betweenness centrality ($BC_n$),  (ii) Local clustering coefficient ($CC_n$), (iii) Closeness centrality ($CLC_n$), (iv) Page rank ($PR_n$) and (v) and Eigenvector centrality ($EC_n$).  
\par {\it (i) Betweenness centrality ($BC_n$)} \cite{newman2003structure,newman2018networks,boccaletti2006complex}. It captures the the relative importance of a node within  a network. This is based on shortest paths: how many times a node is used to find the shortest paths between source and target. Mathematically, the betweenness centrality of a node $n$ can be written as 
\begin{eqnarray}
BC_n&=&\sum_{s,t} \frac{p_{st}(n)}{p_{st}},
\end{eqnarray}
where $p_{st}(n)$ is the number of shortest paths between the patch $s$ and $t$ which passes through the patch $n$ where as $p_{st}$ is the total number of shortest paths between $s$ and $t$.
\par {\it (ii) Local clustering coefficient ($CC_n$)} \cite{newman2003structure,watts1998collective,boccaletti2006complex}. The local clustering coefficient is defined as ratio between  the number of links exist in the neighbors of a node $n$ and all possible links within those neighbors. It signifies how the neighbors are connected to each other. In mathematical expression, for undirected graph, we can write 
\begin{eqnarray}
CC_n&=&\frac{2q}{d_n(d_n-1)},
\end{eqnarray}
where $q$ is the number of actual links exist among neighbours of the node $n$. As the number of nodes (patches) in the neighbours is $d_n$, the maximum number links can exist among the neighbours is $\frac{d_n(d_n-1)}{2}$.

\par {\it (iii) Closeness centrality ($CLC_n$)} \cite{newman2018networks,newman2003structure}.
The closeness centrality of vertex $n$ is the mean shortest path from vertex $n$ to every other vertex in the connected network. Therefore, the nodes which are hubs (central) having shorter distance to others will have smaller centrality value. Therefore, the inverse score of closeness centrality can  capture the linear relation i.e higher numbers will have greater centrality. 
The inverse measure also ensures the central nodes (having large connections) may have good chance to be high $CLC$.

\par {\it (iv) Pagerank ($PR_n$)} \cite{brin2012reprint,litvak2007degree}.
Page rank is practically used to understand the importance of website pages. The ranking can be obtained from a simple iterative process as follows 
\begin{eqnarray}
{\widetilde{PR}}_{t+1}(P_n)=\sum_{P_j\in B_{P_n}}\frac{\widetilde{PR}_t (P_j)}{|P_j|},
\end{eqnarray}
where 
$\widetilde{PR}_{t+1}(P_n)$ 
is the page rank of the page $P_n$ at the iteration  $t+1$ where initial probability distribution is $\frac{1}{M}$. 
$M$
 is the total number of pages. 
$B_{P_n}$ is the 
set of pages directing to $P_n$ and $|P_j|$  is the number of outlinks from page $P_j$. 
{ It is already established  that in-Degree and Page Rank of Web pages are correlated to each other \cite{litvak2007degree}.}

\par {\it (v) Eigenvector centrality ($EC_n$)} \cite{newman2018networks,newman2003structure}.
This is natural extension of existing degree vector. It can be defined as 
\begin{eqnarray}
x_n &=& \frac{1}{\lambda} \sum_{j=1}^M A_{ij} x_j.
\end{eqnarray}
Here $A$ is the adjacency matrix and here we use 
$x_n$ as $EC_n$. 
 Here $\lambda$
 is the eigenvalue of the matrix. According to Perron-Frobenius we may use centrality as the elements of eigenvector having largest eigenvalue \cite{newman2018networks}.
 \begin{figure}
 	{\includegraphics[width= 1\textwidth]{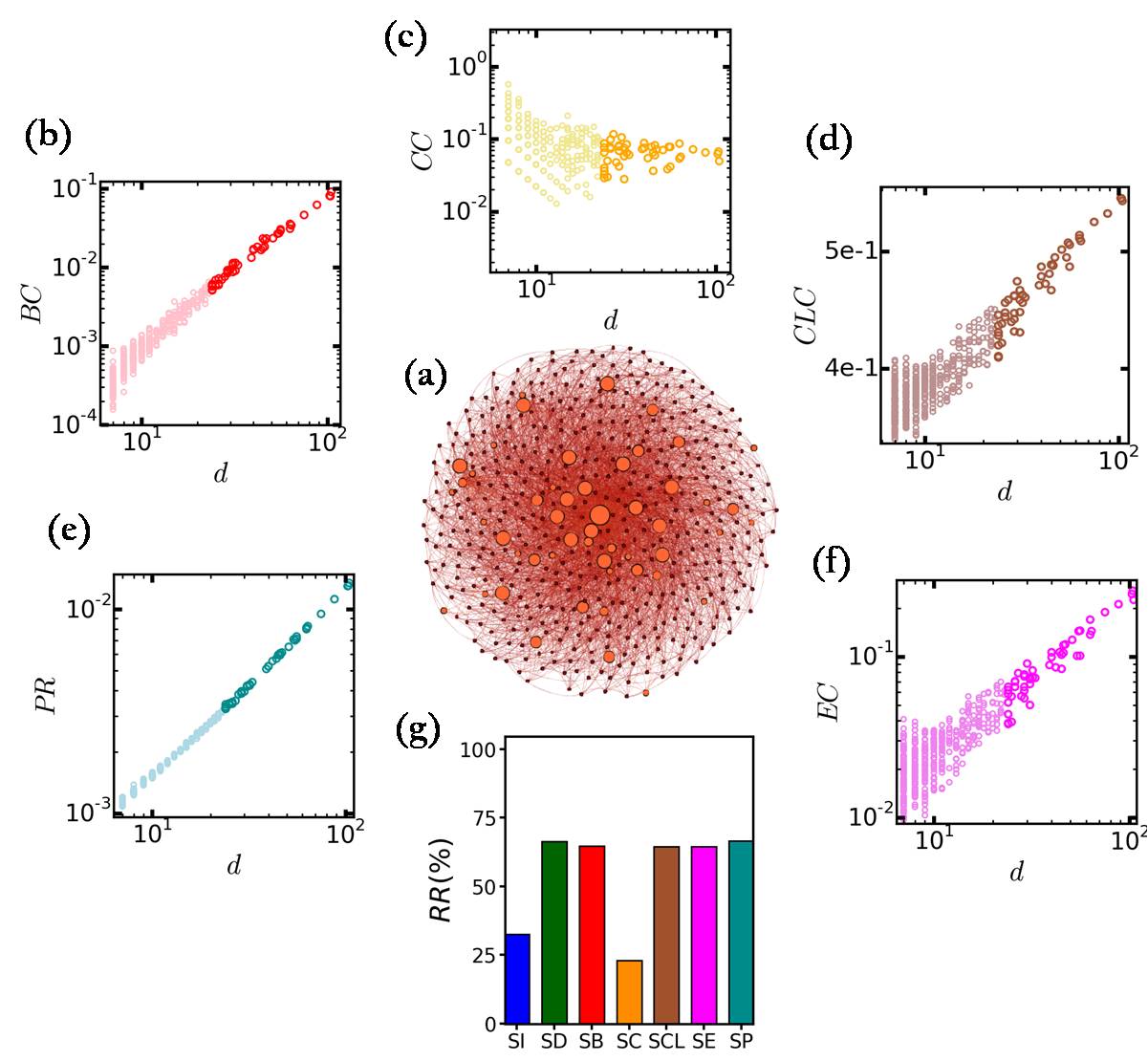}}
 	\caption{\textbf{Centrality measures vs degree ($d$).} (a)The size of the network is $500$. The brick red color in the network represents the  $10\%$  high degree nodes. Degree ($d_n$) vs.\ (b) Node betweenness ($BC_n$), (c) clustering coefficient($CC$), (d) closeness centrality ($CLC_n$),(e) page rank($PR_n$) and (f) eigenvector centrality($EC_n$) of a scale-free network.  Deep colors of each panel represent  $10\%$ of high degree nodes.  (g) Relative reduction of $\mathcal{I}_{\rm max}$ for choosing $10 \%$ nodes having higher $d_n$, $CC_n$, $BC_n$, $CLC_n$, $PR_n$,  and $EC_n$.}
 	\label{fig:sm_4}	
 \end{figure}
Eigenvector centrality determines the importance of each node. The nodes with high number of connections may have large centrality, however, specific nodes with less connection may outrank them.  
Note that, page rank is a variant of eigenvector centrality.\\
We have checked the relation between the degree with other nodal properties  for a network having $500$ nodes having scale-free feature (the exponent of the degree distribution is fixed at $3$). In the network, we have chosen $10\%$  nodes having higher degrees. The network is reported in Fig.\ \ref{fig:sm_4}(a), where high degree nodes are shown in brick red with bigger size. 
At first, we have plotted degree of each node with respect to its betweenness centrality (Fig.\ \ref{fig:sm_4}(b). Clearly, the 
$10\%$ 
high degree nodes (deep red color) are  positively related to the nodes having high betweenness centralities.  The relationship between  closeness centralities and degrees  is shown in Fig.\ \ref{fig:sm_4}(d), where $10\%$ high degree nodes are depicted with brown color. The same features are also validated for  page rank score and eigenvector centrality (Fig.\ \ref{fig:sm_4} (e-f), where
 $10\%$ 
 high degree nodes  are marked with cyan and magenta color respectively). The positive correlation for the  large values of  
$BC_n$, $EC_n$, $CLC_n$ and  $PR_n$  
with nodes having high degrees ensures that test-kit based intervention strategy can successfully mitigate the disease if we apply them on the nodes having higher degree, betweenness, closeness, eigenvector centralities or higher page ranks.
Note that, nodes with higher clustering coefficient cannot capture the degree heterogeneity, therefore unable to reduce infection height as well as final outbreak size. To check the impact of all nodal characteristics, we have 
plotted  (Fig.\ \ref{fig:sm_4}(g)) the relative reduction of $\mathcal{I}_{\rm max}$  by applying test-kits in $10\%$ nodes with higher degree (SD, green) or higher centrality measures such as page rank (SP, dark cyan), closeness (SCL, brown), betweenness (SB, red), eigenvector centrality (SE, magenta)). Clearly,  the nodes with higher scores perform well for all cases except in clustering coefficient. 
\section{Description of real networks}
\label{Description of real networks}
We have checked  (see main text) the impact of   degree based strategy for two real networks: one is a connectivity pattern of international airports  through flights and the other is a transportation network  within  Wards of Kolkata municipality corporation.  Here we give the details of both networks.

\par (i) A global airport network of nodes $1292$  linked
through $38, 377$ directional air-routes \cite{hens2019spatiotemporal,brockmann2013hidden} is considered. 
 
\par (ii) Next,  we construct  transportation connectivity within Kolkata city. The Kolkata municipality has 141 Wards.  We consider them as nodes in the network. At first, we have connected links between the adjacent Wards and generated a lattice like graph i.e we draw a link between Ward
 $A$ and $B$ if they are adjacent to each other.  The small three-wheeler cars (auto) and the pedestrians  naturally move from one Ward to the adjacent Ward reflecting the almost regular connectivity pattern of the lattice. In this setup, the  total number of connections in the lattice are $698$. \\
 Apart from pedestrian mobility,
 we consider the bus connectivity and metro network within  Kolkata city. Maparu {\it et al} showed that in Kolkata,  the bus routes can be subdivided into ten important zones \cite{maparu2010methodology}. Based on   homogeneity in land and population (see Table 1 in the same paper) we  find out the most significant places  of those zones   and connect them to each other (formation of cliques) as there will be bus connections (major bus routes) between them. We have also added  some random links among all other wards to map  several  small  insignificant bus routes.  
Establishing the long distance connection between wards through bus we further add links in the  original lattice to map the metro links between the Wards. As the typical timescale of the movement of people through metro is faster, we have created a complete graph (clique) between the Wards which have metro stations. 
Finally the  network has   $2524$ links and $141$ nodes. 
{ For numerical simulations, we consider the total population in each ward to be 10000. We randomly select two wards and set the initial infection in that wards as 10.}

\bibliography{Kit_bib}
\bibliographystyle{apsrev4-1}



\end{document}